%% file: ms_astroph.tex
\documentclass[12pt,preprint,longabstract]{aastex}

\newcommand{\dss}{$\delta$~Scuti~stars}

\newcommand{\bceph}{$\beta$~Cephei}
\newcommand{\bcephss}{$\beta$~Cephei stars}
\newcommand{\nueri}{$\nu$~Eridani}

\newcommand{\deltam}{\Delta^{-}}
\newcommand{\deltap}{\Delta^{+}}
\newcommand{\edelta}{\epsilon_{\Delta^{+,-}}}
\newcommand{\adelta}{{\epsilon_{A}}}
\newcommand{\vaiss}{Brunt-V\"ais\"ala}

\newcommand{\dst}{\displaystyle}
\newcommand{\muHz} {{\mu\mbox{Hz}}}
\newcommand{\kms} {{\mathrm{km}\,\mathrm{s}^{-1}}}
\newcommand{\msol} {{\mathrm{M}_\odot}}
\newcommand{\rsol} {{\mathrm{R}_\odot}}
\newcommand{\teff} {{T_{\mathrm{eff}}}}
\newcommand{\mmsol} {{M/\msol}}
\newcommand{\rrsol} {{R/\rsol}}
\newcommand{\lteff} {{\log\teff}}

\newcommand{\vsini} {{\mathrm{v}\!\sin\!i}}

\newcommand{\logl} {{\log(L/L_\odot)}}
\newcommand{\amlt} {{\alpha_{\rm MLT}}}
\newcommand{\dov} {{d_{\rm ov}}}
\newcommand{\ml} {{M-Z}}
\newcommand{\xin}{X_{\rm i}}
\newcommand{\xc} {{X_{\rm c}}}
\newcommand{\rc} {{r_{\rm c}}}
\newcommand{\rcr} {{\rc/R}}
\newcommand{\cpd} {{\rm d}^{-1}}

\newcommand{\omegas} {{\omega_{\rm s}}}
\newcommand{\omegac} {{\omega_{\rm c}}}

\newcommand{\graco} {{\sc graco}}
\newcommand{\filou} {{\sc filou}}
\newcommand{\cesam} {{\sc cesam}}
\newcommand{\cles} {{\sc cles}}
\newcommand{\losc} {{\sc losc}}
\newcommand{\wnj} {{\sc warsaw-new jersey}}

\newcommand{\wpack} {{\sc warsaw-new jersey package}}
\newcommand{\gpack} {{\sc granada package}}
\newcommand{\lpack} {{\sc li\`ege package}}

\newcommand{\eqn} [1] {
\begin{equation}
#1
\end{equation}}

\shorttitle{Seismology of the \bceph\ star \nueri}
\shortauthors{Su\'arez et al.}

\begin{document}

\title{Seismology of \bceph\ stars: differentially-rotating models
       for interpreting the oscillation spectrum of \nueri}
       
\author{J.C. Su\'arez\altaffilmark{1,2} and A. Moya\altaffilmark{1}
        and P.J. Amado\altaffilmark{1} and S. Mart\'{\i}n-Ruiz\altaffilmark{1}
        and C. Rodr\'{\i}guez-L\'opez\altaffilmark{3,4,1}
	and R. Garrido\altaffilmark{1}}
	
\altaffiltext{1}{Instituto de Astrof\'{\i}sica de Andaluc\'{\i}a (CSIC),
                Granada, Spain.}
\altaffiltext{2}{LESIA, Observatoire de Paris-Meudon, UMR8109, Meudon,
                 France.}    
\altaffiltext{3}{Laboratoire d'Astrophysique de Toulouse-Tarbes,Universit\'e de
                 Toulouse, CNRS. 31400-Toulouse, France.}
\altaffiltext{4}{Universidade de Vigo, Dpto. de F\'{\i}sica Aplicada, 36310
                 - Vigo, Spain.}

\begin{abstract}
             A method for the asteroseismic analysis of \bceph\ stars
	     is presented and applied to the star \nueri. The method is based
             on the analysis of rotational splittings, and their asymmetries
             using differentially-rotating asteroseismic models.
	     Models with masses around $7.13\,\msol$, and ages around
             14.9\,Myr, were found to fit better 10 of the 14 observed
             frequencies, which were identified as the fundamental radial mode
             and the three $\ell=1$ triplets ${\rm g}_1$, ${\rm p}_1$, 
	     and ${\rm p}_2$. The splittings and aymmetries found for
             these modes recover those provided in the literature, except for
             ${\rm p}_2$. For this last mode, all its non-axysimmetric
	     components are predicted by the models. 
	     Moreover, opposite signs of the observed and predicted splitting
	     asymmetries are found. If identification is confirmed, this
	     can be a very interesting source of information about the
	     internal rotation profile, in particular in the outer 
	     regions of the star.

	     In general, the seismic models which include a description for
             shellular rotation yield slightly better results as compared with
             those given by uniformly-rotating models. Furthermore, we show that
             asymmetries are quite dependent on the overshooting of the
             convective core,  which make the present technique suitable for
             testing the theories describing the angular momentum redistribution
             and chemical mixing due to rotationally-induced turbulence.	
\end{abstract}


\keywords{stars:~evolution --stars: individual: $\nu$~Eridani ---
          stars:~interiors -- stars:~oscillations (including pulsations)
          stars:~rotation --(Stars:~variables:)~$\beta$~Cephei}

\section{Introduction}\label{sec:intro}

The \bceph\ star \nueri\ (\object{HD 29248}) is nowadays one of the most
in-depth
studied stars. Classified as a B2III star, it presents a relatively simple
internal structure, characterised by a large convective core. The $\kappa$
mechanism located in the metal opacity bump situated around
$2\times10^{5}\,\mathrm{K}$ \citep{DziembowskiAlosha93, GautschySaio93}, drives
its oscillations. In addition, the pulsation periods of \bcephss, varying
in the range of 3--8 hours, makes them suitable for detecting and
analysing oscillation frequencies. Several photometric and spectroscopic
multisite observational campaigns for \nueri\ have been set up since 2002 with
subsequent frequency analysis from
spectroscopy \citep{Aerts04nueri} and from photometry 
\citep{Handler04nueri,Jerzy05nueri}. In these later
works, two independent low-frequency, high-order ${\rm g}$ modes were 
detected. As a consequence, \nueri\ can also be classified as a 
SPB star. In terms of light curves, these multisite observations 
constitute the largest time-series ever collected for a \bceph\ star. 
Nowadays, \nueri\ presents the richest oscillation spectrum 
(14 independent frequencies) of the \bceph-type class.
Such privileged scenario for asteroseismology has led to several authors
to analyse its oscillation spectrum and perform seismic models of the star.

Several attempts have been carried out to provide a plausible seismic
model which explains the observed frequencies of \nueri. In
\citet{Ausseloos04} (hereafter ASTA04), a massive exploration of standard and
non-standard stellar models was undertaken in order to fit the oscillation
data. The authors showed that an increase in the relative number fraction
of iron throughout the whole star, or a large decrease in the initial 
hydrogen abundance, made the stellar models satisfy all the observational
constraints, in particular, the modes around the fundamental
radial mode are predicted unstable.

\citet{Alosha04nueri} performed a seismic analysis of the
oscillation spectrum of \nueri\ taking the excitation of modes into account.
In that work only three frequencies were fitted, failing to reproduce
mode excitation in the broad observed frequency range of the ($\ell=1, p_2$)
modes, associated with the highest frequency peak in the spectrum.
Nevertheless, they also inferred some properties of the internal rotation rate
using the rotational splittings of two dipole ($\ell=1$) modes
identified as ${\rm g}_1$ and ${\rm p}_1$. In particular, their results suggest
that the mean rotation rate in the core and the $\mu$-gradient zone is about
three times higher than in the envelope, for their two standard models
fitting the three aforementioned frequencies. 

Recently, \citet{Wojtek08nueri} have analysed the impact of considering 
uncertainties in the opacity and element distributions on the interpretation of
\nueri's oscillation spectrum. No satisfactory explanation of the low-frequency
modes was 
found. Moreover, the authors concluded that some enhancement of the opacity in 
the driving zone is required.

The rotational splitting asymmetries of \nueri\ have also been
studied under different hypothesis. In particular, 
\citet{Dziembowski03nueri} suggested that the asymmetry of the 
$\ell=1$ triplet (around $5.64\,\cpd$), as measured by \citet{vanHoof61},
could be explained by two principal effects: the
quadratic effects of rotation and a strong magnetic dipole field of the
order of 5--10\,kG. Such a magnetic field was searched for by \citet{Schnerr06}
using spectropolarimetry with no success. 

\clearpage
\input{tab1.tex}
\clearpage
Motivated by these results, the present work aims at 
performing a complete modelling of \nueri\ taking 
the effect of rotation up to second order into account, with the
special feature of considering the presence of a radial
differential rotation in the seismic modelling. To do so,
a method based on the analysis of rotational splittings and 
their asymmetries is discussed.

The paper is organised as follows: Section~\ref{sec:modelling}
describes the modelling procedure and provide details of
both the evolutionary models and the oscillation spectra
computation. In Section~\ref{sec:constparam}, the different
sources for constraining the stellar parameters are 
compared, which include a stability analysis.
Then, Sections~\ref{sec:procedure} and \ref{sec:discussion} 
explain the method here presented and discuss its application to
the particular case of \nueri. Finally, conclusions and final
remarks are written in Section~\ref{sec:conclusions}.

\section{Seismic modelling}\label{sec:modelling}

The seismic modelling described below consists in the computation
of evolutionary models and their corresponding adiabatic and
non-adiabatic oscillation spectra. This is described in the
following sections.

\subsection{Equilibrium models}\label{ssec:eqmodels}

To theoretically characterise \nueri, we build equilibrium models representative
of the star with the evolutionary code \cesam\ \citep{Morel97}. In particular,
models taking first-order effects of rotation into account are constructed.
Such models are the so-called pseudo-rotating models, whose spherically averaged
contribution of the centrifugal acceleration is included by means of an
effective gravity ${\rm g}_{\mathrm{eff}}=g-{\cal A}_{c}(r)$, where ${\rm g}$ is
the local gravity, $r$ is the radius, and
${\cal A}_{c}(r)=2/3\,r\,\Omega^2(r)$ is the
centrifugal 
acceleration of matter elements. This spherically averaged component of the
centrifugal 
acceleration does not change the order of the hydrostatic equilibrium equations
\citep{KipWeig90}. The non-spherical components of the centrifugal acceleration
(which are not included in the equilibrium models), are included in the
adiabatic 
oscillation computations (see next section) by means of a linear perturbation
analysis 
according to Soufi, Goupil \& Dziembowski~(1998) (see also Su\'arez, Goupil \&
Morel~2006).
It is possible to evaluate the impact of a differential
rotation, using two simple hypothesis when prescribing the rotation profile
\citep{Sua06rotcel}: 1) instantaneous transport of angular momentum  in the
whole star (global 
conservation) which thus yields a uniform rotation, or 2) local conservation 
of the angular momentum (shellular rotation), except in the convective
core whose rotation is assumed to be rigid. In both cases, no 
mass loss is considered at any evolutionary stage, that is, the total 
angular momentum is assumed to be conserved. These hypothesis represent
extreme cases, so reality is presumably somewhere in between. In fact,
similar rotation profiles have been found when analysing the evolution of
giant stars including rotationally induced mixing of chemical elements
and transport of angular momentum \citep{MaederMeynet04}. 
 
Input physics have been adequately chosen for main sequence B stars.
Particularly, for the mass range treated in this work ($7$--$13\,\msol$), the
CEFF 
equation of state \citep{ceff} is used, in which, the Coulombian correction to
the classical EFF \citep{Eggleton73} has been included. The opacity tables are
taken
from the OPAL package \citep{Igle96}. For the metal mixture, the abundances
given by \citet{GrevesseNoels96} are used.

A weak electronic screening is assumed, which is valid in the evolutionary
stages considered in this work 
\citep[see][ for more details]{Clayton68}. For adiabatic stellar models, 
the Eddington's~$T(\tau)$ law (grey approximation) is considered for the
atmosphere reconstruction. For non-adiabatic stellar models, the atmospheres
are reconstructed using the Kurucz equilibrium atmospheric models
\citep{Kurucz93cd13} from a specific Rosseland optical depth until the
the last edge (around $\tau=10^{-3}$) of the star is reached.

Convection is treated with the mixing-length theory \citep{BohmVit58} 
which is parametrised with $\alpha_{\rm MLT}=l/H_{\rm p}$, 
where $l$ is the mean path length of the convective elements, and $H_{\rm p}$
is the pressure scale height.
In addition, we use the overshoot parameter, defined as 
$\dov$ ($l_{\rm ov}$ being the penetration length of
the convective elements).

The grid of equilibrium models has been constructed with steps of 0.002 dex,
0.01, and 0.05, in the metallicity $Z$, overshoot parameter $\dov$, and 
mass $M$, respectively. 
\clearpage
\begin{figure}
     \plotone{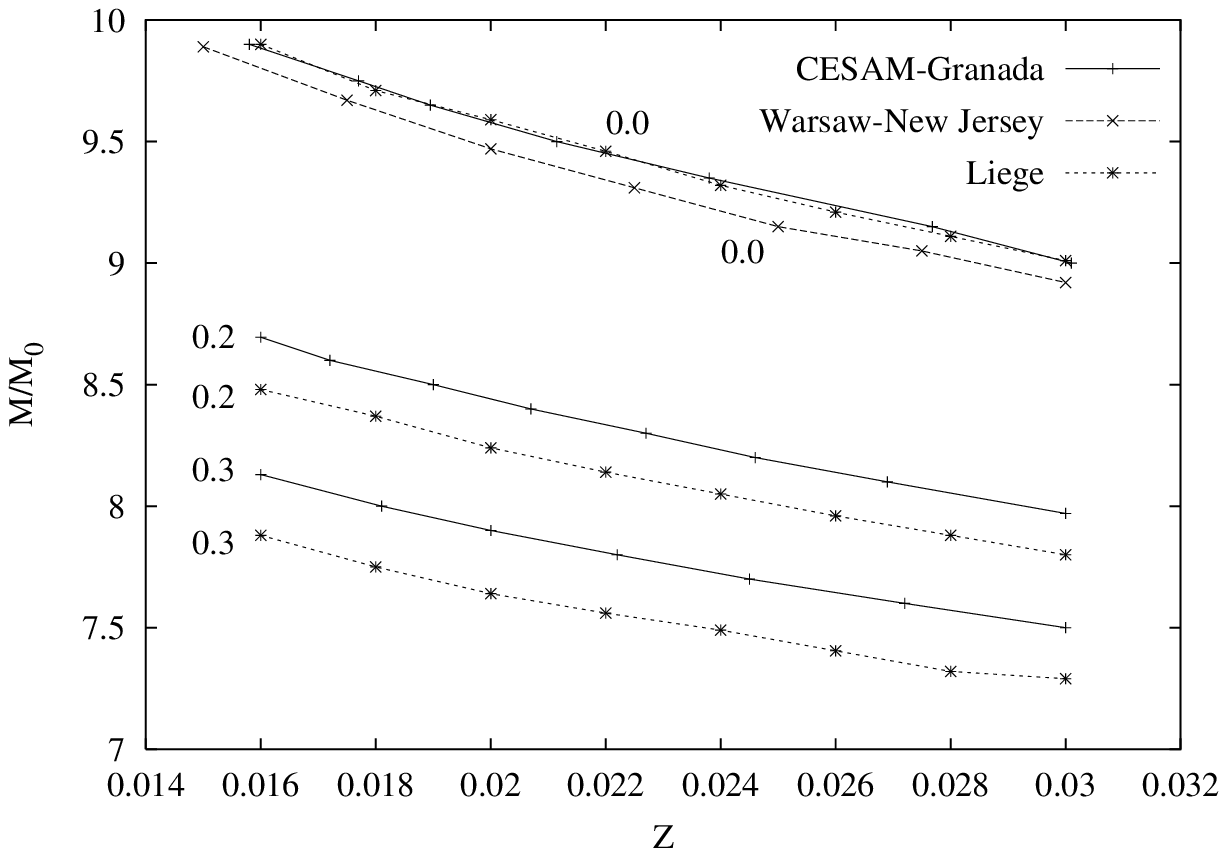}     
     \caption{Mass-Metallicity relation fitting the observed frequencies 
              $f_1$ and $f_4$ obtained from three different codes: \cles, 
               \gpack\ and \wnj\ codes. Two comparisons are shown: 1) for 
	       models without overshooting (\lpack, \gpack\ \& \wnj) and 2) for 
	       models with $\dov=$0.2 and 0.3 (\lpack, \gpack).}  
	       \label{fig:compcodes}
\end{figure}
\clearpage

\subsection{Oscillations}\label{ssec:oscillations}

Adiabatic eigenfrequencies of selected pseudo-rotating models described in the
 previous section have been computed with the oscillation code \filou\ 
\citep[see][]{filou,SuaThesis}. In this code, oscillation frequencies
are obtained by means of a perturbative method taking into account up to 
second order effects of rotation. In the 
case of near-degenerate frequencies, i.e. when two or more frequencies are close
to 
each other ($\omega_{nlm}\sim\omega_{n'l'm}$), the corrections for near
degeneracy
are included. As detailed in \citet{Sua06rotcel},
the oscillations computation also takes into account the presence of
a radial differential rotation profile under the form
\eqn{\Omega(r)={\bar \Omega}\,\Big(1+\eta_0(r)\Big)\label{eq:defOmega}}
where ${\bar \Omega}$ represents the angular rotational velocity at 
the surface and $\eta_0(r)$ a radial function. This rotation 
profile is equivalent to the \emph{shellular} rotation profile 
obtained with the pseudo-rotating models described in the previous
section.

Concerning the instability computations, non-adiabatic theoretical 
observables are obtained using \graco\ code \citep{Moya04}. 
This code solves the stellar pulsation equation in a non-adiabatic non-rotating 
frame by dividing the star in two parts, 1) the
 interior, by means of the non-adiabatic equations described in \citet{Unno89},
 and 2) the atmosphere, taking into account the interaction with the pulsation 
as prescribed by \citet{Dupret02}. 

\subsection{Comparison of numerical seismic packages}\label{ssec:mlrelation}

As it is widely known, in stellar modelling, and in particular, 
in the field of stellar seismology, the use of different codes 
with different numerical techniques, can be crucial for the correct
interpretation 
of seismic data \citep[see][]{Moya08estacorot}. 
ASTA04 compared two seismic packages: the \lpack, which comprises
the evolutionary code \cles\ \citep{Scuflaire07}, plus the oscillation
code \losc\ \citep{Boury75cles}, and the \wpack\ 
\citep[by][]{Dziembowski03nueri} which comprises the \wnj\ code 
and its corresponding oscillation code. Following ASTA04, 
these codes are compared (through a mass-metallicity relation) with our seismic 
modelling package, the \gpack, which is composed
by the evolutionary code \cesam\ \citep{Morel97}, and the oscillation code
\graco\  \citep{Moya08graco} and \filou\ \citep{Sua08filou},
described in Sections~\ref{ssec:eqmodels} and \ref{ssec:oscillations},
respectively. Hereafter, those three packages are called LP, WP, and GP,
respectively.

Similarly as done in ASTA04, the mass $M$, 
 metallicity $Z$, overshoot parameter $\dov$, and 
 initial hydrogen content $\xin$, are then varied such as to fit the two
observed frequencies, $f_1$ and $f_4$ (Table~\ref{tab:freqobs}) by using 
non-rotating models. This procedure yields a mass-metallicity relation for each
$\dov$ value, which is compatible with the similar study described in
ASTA04 (see Fig.~\ref{fig:compcodes}).
It is found that, under similar conditions, i.e. for a given metallicity,
$\amlt$ and $\dov$, GP predicts higher mass values, around $0.25\,\msol$.
These differences increase slightly
when $\dov$ increases. Such a different behaviour can be explained by the
different treatments of the overshooting implemented in the evolutionary codes.
Indeed, for the overshooting description, the Li\`ege 
evolutionary code only takes the density variations (in the overshooted region)
into account, which slightly affects the temperature gradient, whereas \cesam\
considers an additional restriction by imposing that the \emph{real} temperature
gradient must be equal to the adiabatic one, i.e. $\nabla=\nabla_{\rm ad}$.
This would imply that, either transport of heat is purely radiative,
or it is efficiently transported outwards from the stellar core through
convective movements (as a result of the overshooting), respectively.
This constitutes an interesting challenge for asteroseismology 
because the oscillation modes are sensitive
to the physical description of the $\mu$-gradient zone.

\clearpage
\begin{figure}
     \plotone{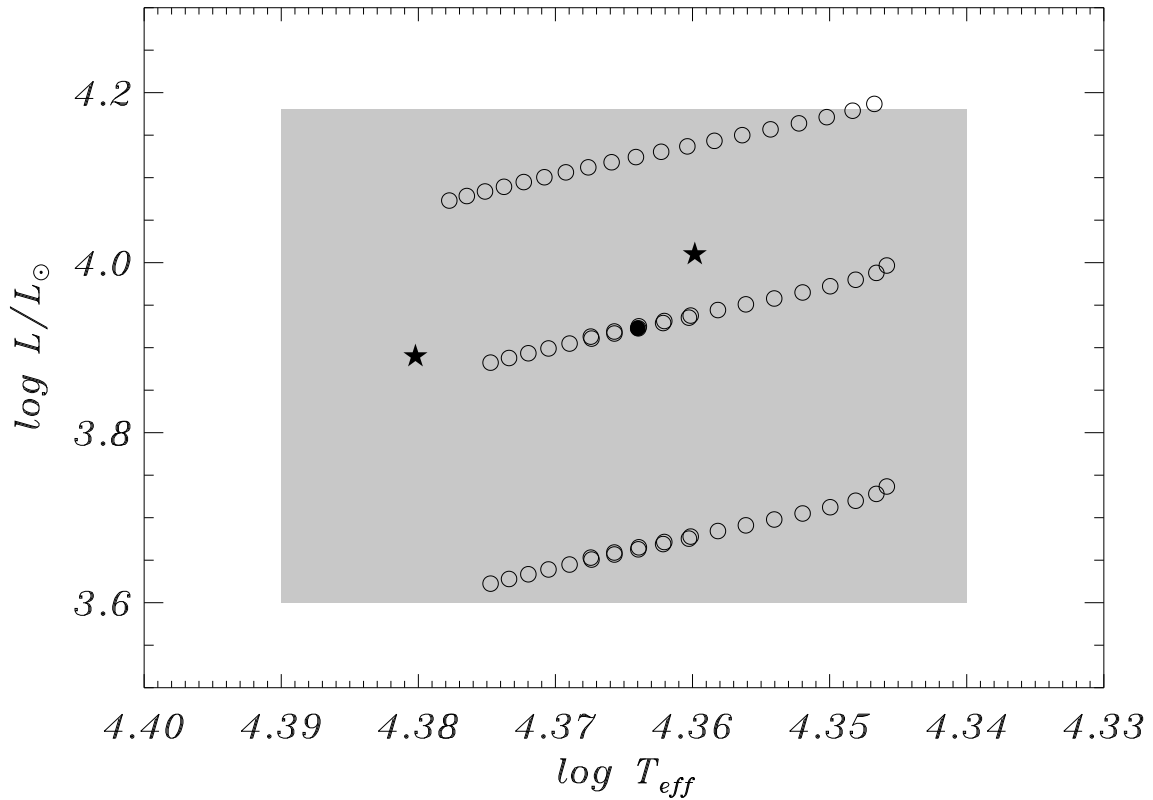}
     \caption{HR diagram containing the two estimates for the effective
             temperature and luminosity (filled star symbols) adopted in this
             work for \nueri\ (more details in Section~\ref{ssec:HR}), with the
             corresponding photometric uncertainty boxes (shaded area). For
             illustration, three evolutionary tracks of 7.60, 7.13 and
             $7\,\msol$, are also depicted. Those models have been computed
             using an initial hydrogen content of $\xin=0.5$.
	     The filled circle represents the selected model NR1, which
	     predicts unstable and better fits the observed frequencies
             $f_1$, $f_4$, $f_6$, and $f_9$. More details
             Section~\ref{ssec:instab}.} \label{fig:HR}
\end{figure}
\clearpage

\section{Constraining stellar parameters}\label{sec:constparam}

The modelling of any star entails the constraint
of its stellar parameters. For doing so, the choice of the 
set of free parameters to be fixed generally depends on the 
observational material available. In the case of \nueri,
the following space of free parameters was chosen
\eqn{{\cal P}={\cal P}\,(M,t,\dov,Z;\Omega)\,\nonumber\label{eq:param}}
where $M$ is the stellar mass, $t$ the age,
$\dov$ the overshooting parameter, $Z$ the metallicity,
and $\Omega$ the angular rotational velocity. From photometric and
spectroscopic observations, we search for an estimate of the
star location in the HR diagram. This allows us to constrain
the metallicity and to have an estimate of the evolutionary
stage of the star. Stability analysis significantly reduces the region of 
the HR diagram in which representative models can be searched.

Then, the fitting of four 
of the observed frequencies permits to better constrain the mass, 
metallicity, evolutionary stage (age) and overshooting parameter
of the star. The small observed $\vsini$ of \nueri\ allows
the use of non-rotating models for this exercise. However, further model
constraining
makes it necessary to take the stellar rotation into account.
In particular, seismic models including shellular rotation
profiles are used (see Section~\ref{sec:modelling}). Finally, analysis
of the rotationally-split modes and their asymmetries allows
to make a refined search for representative models of the star. 

\subsection{Locating \nueri\ in the HR diagram}\label{ssec:HR}

In order to locate the star in the HR diagram we followed the works by
\citet{MorelT06} and \citet{DeRidder04}, based on high-precision
spectroscopy. The error box shown in Fig.~\ref{fig:HR} takes into 
account the results reported in both papers.

%
%
\clearpage
\begin{figure}
  \begin{center}
    \includegraphics[width=8.5cm]{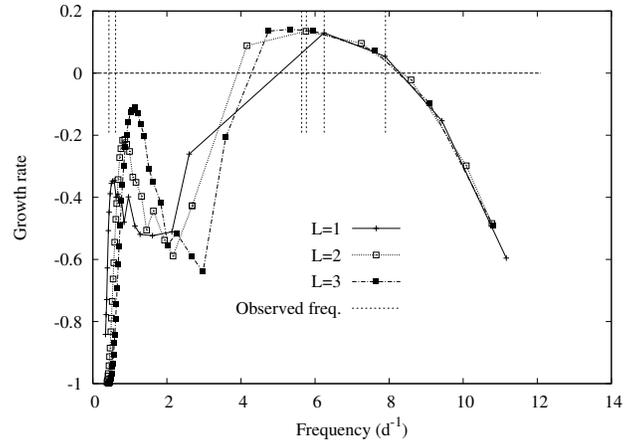}
    \caption{Growth rates $\eta$ as a function of the oscillation frequency.
             Each curve represents the growth rate obtained for oscillation
             modes with different degree ($\ell$ ranging from 0 to 3).
	     The lowest frequencies correspond to the SPB-type pulsation
	     frequencies. Only the frequencies around the fundamental radial 
	     mode are predicted unstable (positive growth rate), which
	     are those kept for the present investigation.}
 \label{fig:growthrates}
  \end{center} 
\end{figure}
\clearpage
On the other hand, the location of stars in the HR diagram depends on both the
rotational 
velocity of the star and the inclination angle with respect to the observer, 
 $i$. A technique to estimate and correct for the effect of fast rotation on
the determination of fundamental
parameters 
for pulsating stars is described in \citet{MiHer99}. This technique was then
refined by \citet{Pe99},
who applied it to \dss\ in clusters. A first consequence that can be extracted
from those works is that rotation increases the size of the 
uncertainty box of stars in the HR diagram (when obtained from photometry). In
general, such additional
uncertainties increase for increasing rotational velocities (for a given
angle of inclination of the star). \textbf{For \nueri, a projected rotational
velocity of $\vsini\sim16\,\kms$ has been observed. This value has been 
derived by \citet{MorelT06} from high-resolution spectroscopy,
taking into account the broadening due to oscillations.  
}
In principle, it can be considered as a slow
rotator so that the additional uncertainties coming from the effect of rotation
can be considered within the already large photometric uncertainties.
\clearpage
\input{tab2.tex}
\clearpage

\subsection{Instability predictions}\label{ssec:instab}

Once the star is located in the HR diagram, models which predict the observed
frequencies to be unstable (considering the constraints on physical parameters
given above) are searched for. 

To do so, no specific \emph{ad-hoc} modification of 
the iron mixture throughout the star is proposed. The search for the best models
is then enhanced by using the $\ml$ relations presented in
Section~\ref{ssec:mlrelation}.
We recall that lines in Fig.~\ref{fig:compcodes} represent the models
which fit simultaneously both $f_1$ and $f_4$. These models constitute
a first guess to the best solution. Finally, our best solution is given
by those models that also fit $f_6$ and $f_9$. 
Consistently with ASTA04, when using standard solar mixture
for the metallicity, an initial hydrogen content of $\xin\sim0.50$ 
is required in order to predict unstable the observed frequencies.
In Fig.~\ref{fig:HR}, the different evolutionary tracks shown have been computed
using $\xin=0.50$ in the ZAMS. From our grid of models (see
Section~\ref{ssec:eqmodels}), the \emph{best} model (filled circle in
Fig.~\ref{fig:HR}) was selected to be the one which predicts unstable and better
fits the observed frequencies $f_1$, $f_4$, $f_6$, and $f_9$. This \emph{best}
non-rotating model is characterised by a mass of $7.13\,\msol$, a solar metallic
mixture (with an initial hydrogen content of $\xin=0.50$), and by the physical
parameters $\amlt=1$, $d_{\mathrm{ov}}=0.28$ (NR1 in Table~\ref{tab:models}).
Note that rotation has not been taken into account in the instability
predictions described above. The lack of theories describing the effect of
rotation on mode stability makes it difficult to estimate this effect for
\nueri. Nevertheless, as stated by \citet{Alosha75}, mode
stability depends predominantly on the effective temperature of the models
\citep[see][for an interesting discussion on this issue]{Sua07gammes}, whose
variations due to rotation are expected to be small (due to the low
rotational velocity of the star).

The results of the stability analysis obtained for that model are depicted in
Fig.~\ref{fig:growthrates}, in terms of growth rate as a function of the
oscillation frequency. As expected, in the frequency region around the
fundamental radial mode, the results obtained for the different spherical
degrees $\ell$ are quite similar, which implies that this parameter cannot be
discriminated. 

\section{Procedure}\label{sec:procedure}

As discussed in Sections~\ref{ssec:eqmodels}-\ref{ssec:oscillations}, the small
rotational velocity of the star makes it plausible to initially adopt the
parameters of the non-rotating models which were found to be representative of
the star. \textbf{In particular, in order to work with models located in the HR
error box which predict unstable the observed frequencies, a value of
$\xin=0.5$ for the initial hydrogen fraction is kept (see previous
section). Indeed, such a value is rather unrealistic. This can be solved
either by considering an ad-hoc iron enhancement in the driving zone
\citep[as done by][]{Alosha04nueri}, or by modifying  (uniformly throughout 
the stellar interior) the relative number fraction of iron (as done by ASTA04).
However, the modelling techniques used here do not allow to make such
modifications. Instead, following ASTA04 we simulate the metallicity change by
modifying the initial hydrogen fraction.
This yields, similarly to ASTA04 (see Table~1 in that paper), models around
$7\,\msol$ and $Z=0.018-0.019$. Similarly to previous works, standard models,
i.e. those with $\xin\sim0.7$, within the HR error box present masses higher
than $8\,\msol$, $Z\sim0.015$, for similar surface rotational velocities. Such
variations with respect to the non-standard models have no
significant impact neither on the rotation profile nor on the splitting
asymmetries.  
}

A systematic search for representative models within the error box was then
performed locally varying
the mass, initial rotational velocity, and age
of the models. The mass was varied around $7.10\,\msol$, in particular from 7 to
$7.20\,\msol$, 
in steps of $0.01\,\msol$. Rotation was considered under two main assumptions:  
uniform rotation and differential rotation 
(see Section~\ref{sec:modelling} for more details). For both assumptions, the
rotational
velocities considered range from to 5 to $20\,\kms$ at the stellar surface.
Then, the modelling was refined by analysing the rotational splittings and
their asymmetries. Rotational splittings are defined as
\eqn{S=\frac{1}{2}\big(\nu_{+1}-\nu_{-1}\big)\label{eq:defs}}
and asymmetries of these splittings are defined as
\eqn{A = \nu_{-1}+\nu_{+1}-2\nu_{0}\,.\label{eq:defasym1}}
where sub-indices $\pm1$ represent the value of the azimuthal 
order $m$ for a given $\ell$. The information provided by
the asymmetries can be completed by the semi-splittings, 
which corresponds with 
$\deltap = \nu_{+1}-\nu_{0}$, and $\deltam= \nu_{0}-\nu_{-1}$, 
respectively.
Both quantities are related such as Eq.~\ref{eq:defasym1} 
transforms into
\eqn{A = \deltap-\deltam\,.\label{eq:defasym2}}
Moreover, this analysis also permits to extract
information about the internal rotation profile of the star,
since, as we discuss in the next sections, 
$S$ and $A$ are both sensitive to changes in the rotation profile.

\section{Results and discussion}\label{sec:discussion}

Models with a mean density of about $\bar \rho=0.064\, {\rm g\,cm}^{-3}$ 
were found to fit better the observed frequencies. 
The models within $\pm0.05\,\msol$ of the mass value calibrated using
non-rotating models yield similar results, but for slightly
different ages. Due to the so small rotational velocity, it is plausible to
assume the same $\amlt$ and $\dov$ parameters which were calibrated by the 
non-rotating models. Therefore, only the mass, age, and
initial rotational velocity of the models were varied.

\begin{figure}
  \begin{center}
    \includegraphics[width=10cm]{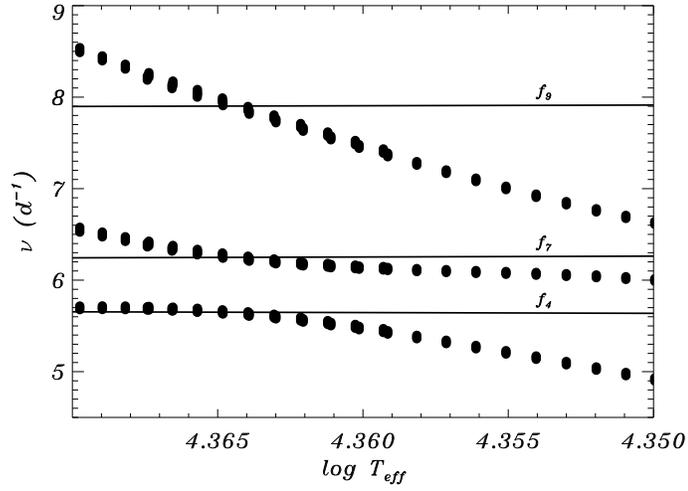}
    \caption{Evolution of the theoretical oscillation frequencies corresponding,
             from to bottom, to the $\ell=1$ triplets ${\rm p}_2$, ${\rm p}_1$
             and ${\rm g}_1$, for the selected $7.13\,\msol$ model. Horizontal
             lines represent the observed frequencies, bottom to top, $f_4$,
             $f_6$, and $f_9$, identified as the $m=0$ components of the
             triplets ${\rm p}_2$, ${\rm p}_1$ and ${\rm g}_1$, respectively.}
             \label{fig:evolfreq}
  \end{center} 
\end{figure}

Models were then selected to fit at least the observed frequencies
$f_1$ (identified as the fundamental radial mode) and the triplet 
$(f_3,f_4,f_2)$, identified as ${\rm g}_1$. With these criteria,
a model with a mass of $7.13\,\msol$, a rotational velocity
(in the surface) of about $7\,\kms$, and an age of about 14.9 Myr
(which corresponds with $\log\teff=4.365$) was found
to better fit the observational frequencies. 
As expected, when rotation
is taken into account, the stellar parameters of the models are similar 
to those of the non-rotating \emph{best} model (NR1). 
For each type of rotation, i.e. uniform and shellular rotation, the 
models better matching the observed frequencies, splittings,
and asymmetries, are UR1 and SR1, whose characteristics 
are summarised in Table~\ref{tab:models} (their corresponding list of 
frequencies are reported in Table~\ref{tab:freqteor}),
\textbf{and whose internal rotation profiles are depicted in
Fig.~\ref{fig:profiles}}.
Figure~\ref{fig:evolfreq} shows the evolution of the theoretical
frequencies of the ${\rm g}_1$, ${\rm p}_1$ and ${\rm p}_2$ $\ell=1$ triplets
for the selected model (assuming shellular rotation).

With these models, nine of the observed frequencies were
identified as two $\ell=1$ triplets: a ${\rm g}_1$ triplet
composed by the observed frequencies $(f_3,f_4,f_2)$, 
a ${\rm p}_1$ triplet which corresponds with $(f_{12},f_6,f_{7})$,
and a ${\rm p}_2$ triplet which corresponds with 
$(f_5,f_9,f_{10})$. Furthermore, $f_1$ was identified as the fundamental radial
mode. \textbf{For the remaining frequencies ($f_{8}$ and
$f_{11}$), the selected models present similar predictions. While $f_{8}$ is
identified as a ($n=0,\ell=2$), all the models match $f_{11}$ with a $\ell\geq7$
mode. In particular the best identification found corresponds to
$(n,\ell,m)=(-3,9,8)$.} 

\textbf{In general, these results are compatible with previous
studies that can be found in the literature, except that we do identify and use
(for asteroseismic purposes) all the non-axisymmetric components of the $\ell=1$
triplets, in particular ${\rm p}_2$. Differences with \citet{Wojtek08nueri}'s
work are principally related to this last split mode, whose identification
depends strongly on the use of standard or non-standard models.}

\clearpage
\input{tab3.tex}
\clearpage
\subsection{Analysis of rotational splittings and asymmetries}\label{ssec:asym}

In Fig.~\ref{fig:deltapmasym}, the predicted
semi-splittings (left column) and asymmetries (right column),
are compared with the corresponding observed values
given in Table~\ref{tab:identif}. For all panels in that figure, 
the shaded vertical regions indicate the range of $\teff$
values given by the best models SR1 and UR1 (see Table~\ref{tab:models}).

This comparison is performed simultaneously for models evolved assuming
uniform and shellular rotation profiles (hereafter UR and SR
models). For the sake of brevity, in the following, 
$A_{x}$, $A_{x}^{\rm o}$, are used to represent the
predicted and observed asymmetries, $x$ representing 
the triplets ${\rm g}_1$, ${\rm p}_1$, or ${\rm p}_2$. Similarly, 
$\Delta^{+,-}$ and $\Delta_{\rm o}^{+,-}$ are used
to represent the predicted and observed semi-splittings.
Let us now examine each selected triplet separately.

\subsubsection{The ${\rm g}_1$ triplet}\label{sssec:g1}

Analysis of the ${\rm g}_1$ triplet
(Fig.~\ref{fig:deltapmasym}, left top panel) reveals that
for the SR models, the predicted $|\deltap|$ remain lower than 
$|\deltam|$ for almost the whole range of effective temperature studied
(almost identical in the shaded region). On the other hand, 
for UR models, $|\deltam|$ are predicted
lower than $|\deltap|$ for effective temperatures higher
than 23014 K ($\log\teff=4.362$), approximately.
The deviation of the semi-splitting
predictions (in the shaded region), defined as 
$\edelta=\Delta^{+,-}-\Delta^{+,-}_{\rm o}$,
is about $2\times10^{-3}\,\cpd$ for the SR models and 
3-$4\times10^{-3}\,\cpd$ for the UR models. 
On the other
hand, cooler SR models would fit better
the observations ($\edelta=10^{-6}\,\cpd$ for models
around $\log\teff=4.35$). 
In general, the evolution of the semi-splittings for
both types of rotation is found to be quite different. This is
somehow expected since ${\rm g}$ modes
are very sensitive to variations of the rotational
velocity of the core. In fact, it was shown by \citet{Sua06rotcel}
that a shellular rotation profile modifies significantly
the radial displacement eigenfunctions, 
especially for ${\rm g}$ and mixed modes. This implies that the 
presence of shellular rotation may affect both the rotational 
splitting itself and its asymmetry. 
The general definition of the first-order rotational splitting kernel
can be written as

\begin{figure}
  \begin{center}
    \includegraphics[width=10cm]{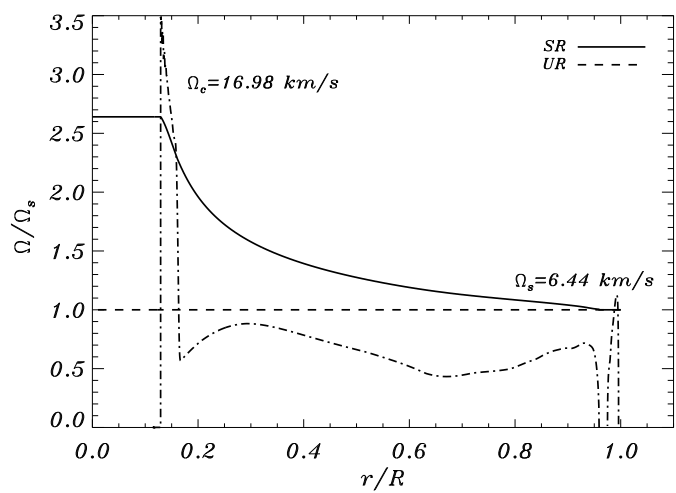}
    \caption{Rotation profile normalised to the rotation velocity in the
             surface, $\Omega_{\rm s}$, as a function of the normalised
	     stellar radius, for the \emph{best} SR model. For comparison,
	     the rotation profile of UR model with similar $\Omega_{\rm s}$
	     is also depicted. For illustration purposes, the \vaiss\ frequency
	     for the SR model (re-scaled as to fit the plot) has been used
	     to indicate the location of the $\mu$-gradient zone
	     ($r\sim0.15-0.2\,\rsol$).}
	     \label{fig:profiles}
  \end{center} 
\end{figure}

\eqn{{\cal K} =\dst{{\Big[\big(2\,y_{01}z_{0}+z_{0}^2\big)+
               \eta_0\big(y_{01}^2+
	       \Lambda z_{0}^2-2\,y_{01}z_{0}-z_{0}^2\big)\Big]
	       \,\rho_0\,r^4}\over
	       {{\dst\int_0^R [y_{01}^2+\Lambda\,z_0^2]\,\rho_0\,r^4
	       \,{\rm d}r}}},
	       \label{eq:defK}}
where $\Lambda=\ell(\ell+1)$, and $y$ and $z$ represent the vertical and
horizontal
displacement normalised eigenfunctions, respectively. The second term within
the square brackets accounts for the presence of shellular rotation
through $\eta_0$, defined in Eq.~\ref{eq:defOmega}. When
a uniform rotation is considered, this term becomes null.
In Fig.~\ref{fig:kernels} such kernels are depicted for the ${\rm g}_1$, ${\rm
p}_1$, and ${\rm p}_2$
triplets. Notice that, when considering a shellular rotation profile, 
a bump in the energy distribution near the $\mu$-gradient zone
(see Fig.~\ref{fig:profiles}) comes up, for ${\rm g}_1$, at the expense of the
energy 
of the outer layers, which could explain
the different behaviour of the semi-splittings predicted by 
the UR and SR models.
\clearpage
\input{tab4.tex}
\clearpage

Concerning the asymmetries, $A_{{\rm g}_1}$ are predicted to diminish for
decreasing
effective temperature (Fig.~\ref{fig:deltapmasym}, right panel on top).
In other words, asymmetries decrease while the star evolves. 
For UR models, such a decreasing is more rapid than
for the SR models, and the asymmetries fit the observed
value at $\log\teff\sim4.357$, which represents a
difference of $\sim400\,$ K with respect to the models
that fit the observed frequencies (shaded
area). On the other hand, the asymmetries predicted by the SR models
never fit exactly the observed value. 
In the shaded region, the deviation of the asymmetry 
defined as $\adelta=A_{{\rm g}_1}-A_{{\rm g}_1}^{\rm o}$, is found to be 
$6\times10^{-4}\,\cpd$ for the SR models whereas for
UR models $\adelta=10^{-3}\,\cpd$. Such deviations
are respectively, two and three orders of magnitude larger than 
the observed asymmetry uncertainty.

As for the ${\rm g}_1$ semi-splittings,
better results
are found for cooler models, especially for UR models around
$\log\teff=4.357$, and for SR models near $\log\teff=4.35$.
In any case, for low order ${\rm g}$ and ${\rm p}$ modes, asymmetries
are sensitive to variations of the rotation
profile near the core. Indeed, as shown
by \citet{Sua06rotcel}, the analytical form of $A$ (using
a perturbative theory) can be written as the second-order term
\eqn{A = -{{6 m^2}\over{4\Lambda-3}}{{{\bar \Omega}^2}\over{\omega_0}}  
     {\cal J}_c\,.}
where ${\bar \Omega}$ is the rotational velocity
at the stellar surface, and $\omega_0$ is the unperturbed oscillation
frequency. The ${\cal J}_c$ integral contains a complex combination of
structure and oscillation terms 
which are modified by the rotation profile and its 
derivatives. Analysis of this term 
is definitely necessary to construct simplified kernels for $A$
(work in progress).
Such kernels will help us to better understand the behaviour 
of the asymmetries, and, especially their sensitivity to
variations of the internal rotation profile.
\clearpage
\clearpage

\subsubsection{The ${\rm p}_1$ triplet}\label{sssec:p1} 

For ${\rm p}_1$, the predicted curves of $|\deltap|$ remain lower than 
those of $|\deltam|$ in the whole range of effective temperature
studied, which is compatible with the observations 
(Fig.~\ref{fig:deltapmasym}, left middle panel). Notice that,
in the shaded region, the SR models predict absolute values for the
semi-splittings
closer to the observed ones ($\edelta\sim10^{-4}\,\cpd$) than 
those predicted by the UR models, for which 
$\edelta=4\times10^{-3}\,\cpd$. However, this situation
is reversed for the splitting asymmetries 
(Fig.~\ref{fig:deltapmasym}, right middle panel).
In particular, around $\log\teff=4.365$, the asymmetries predicted 
by the UR models ($\adelta=6\times10^{-4}\,\cpd$) are slightly closer to the 
observed values than those predicted by the SR models
($\adelta=2\times10^{-4}\,\cpd$). Even so, such differences are 
of the order of magnitude of the observational uncertainty
of $A_{{\rm p}_1}$, which makes it difficult to discriminate between
both types of rotation. Contrary to the ${\rm g}_1$ results, cooler
models do not fit better the observations.

\subsubsection{The ${\rm p}_2$ triplet}\label{sssec:p2} 

\begin{figure*}
  \begin{center}
   
    \includegraphics[width=6.5cm]{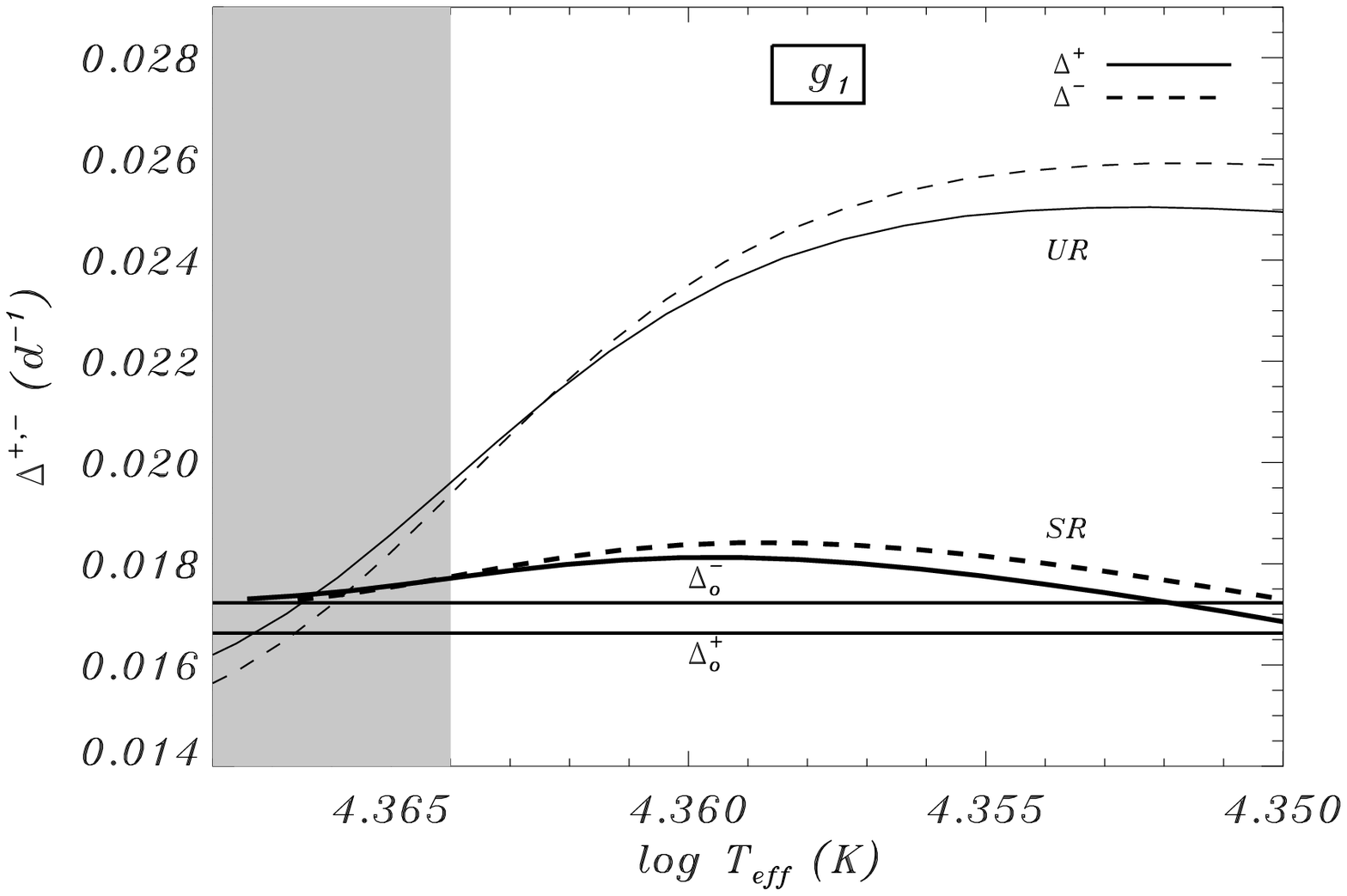}
    \includegraphics[width=6.5cm]{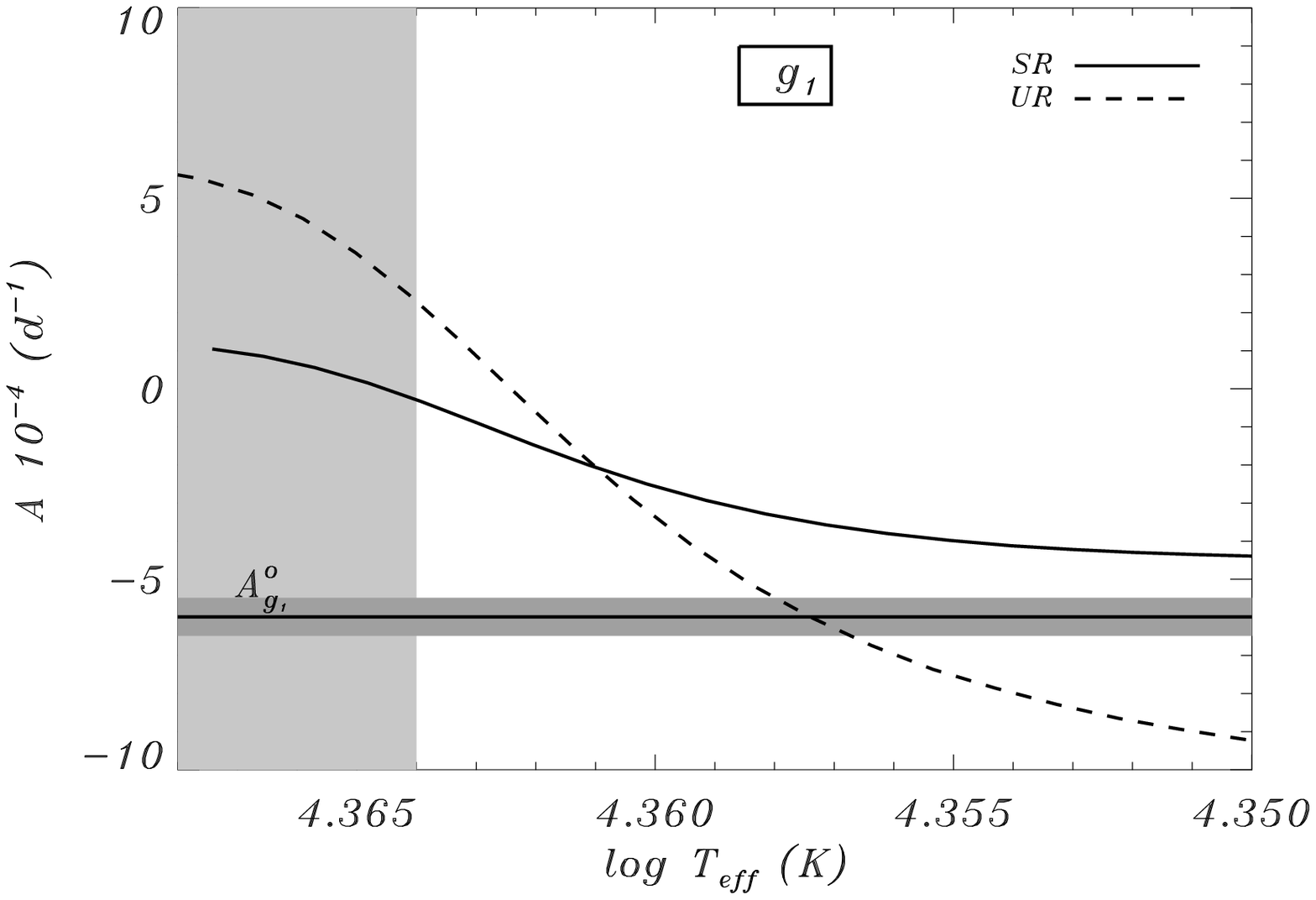}
   
    \includegraphics[width=6.5cm]{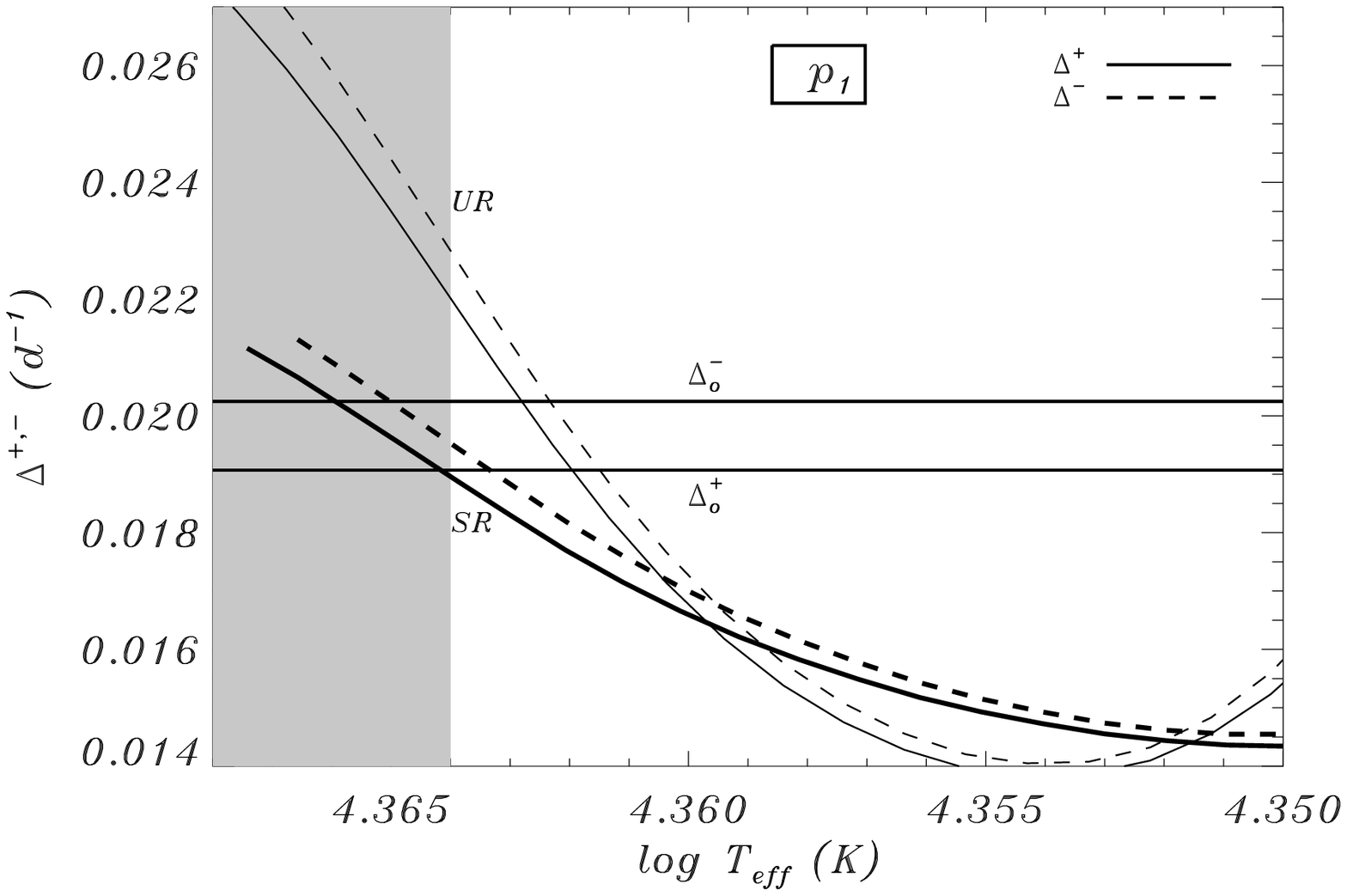} 
    \includegraphics[width=6.5cm]{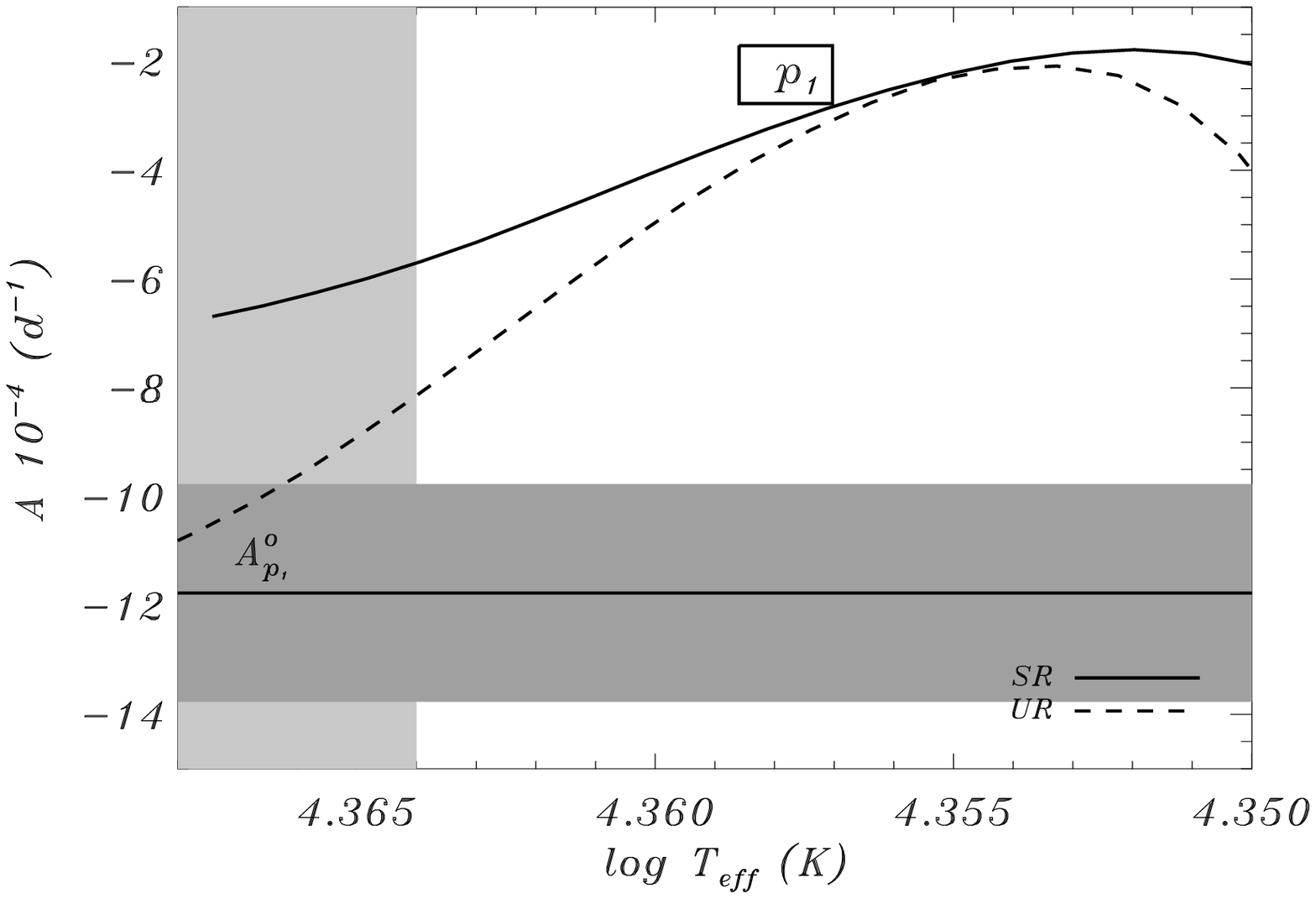} 
   
    \includegraphics[width=6.5cm]{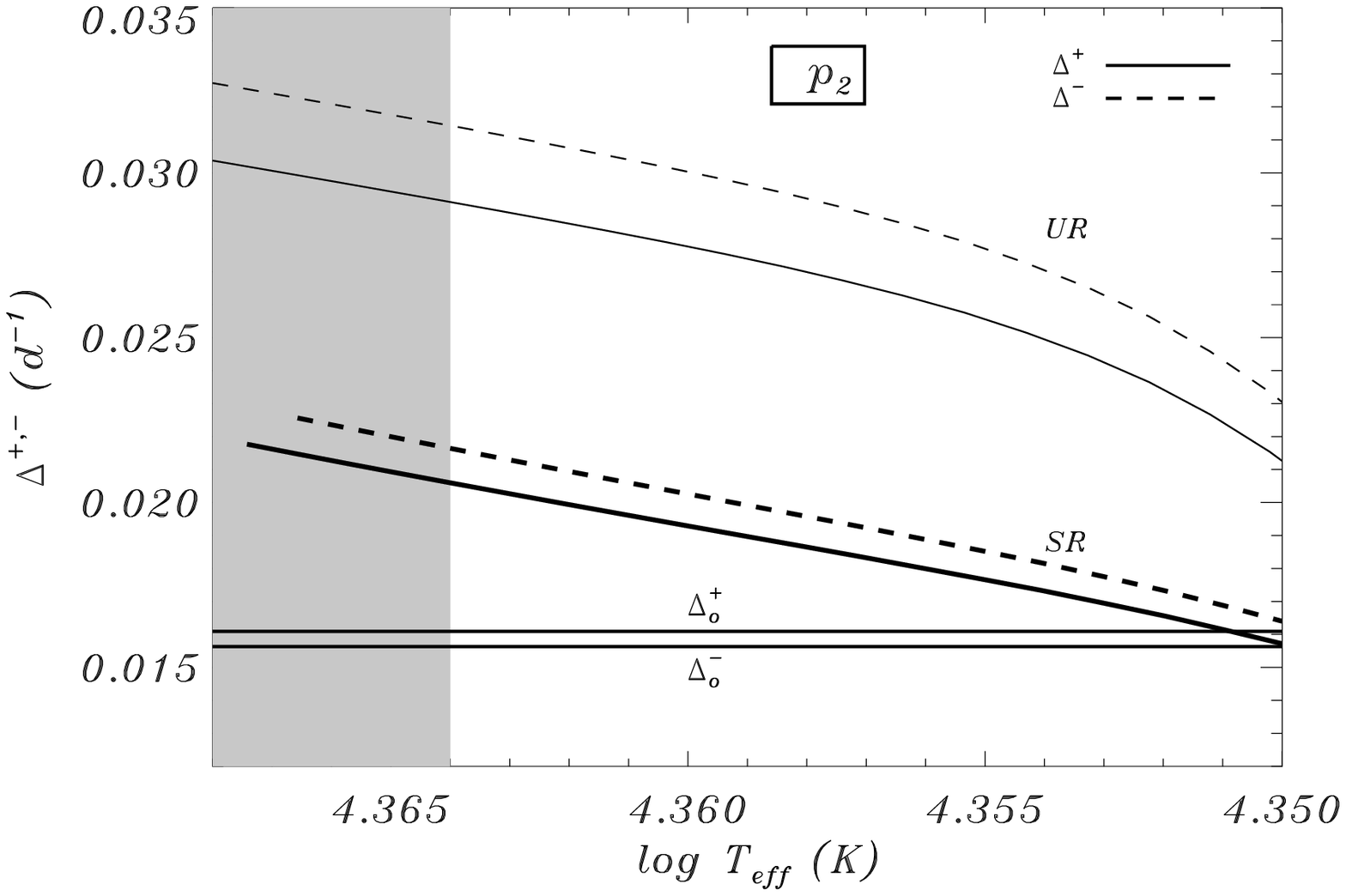} 
    \includegraphics[width=6.5cm]{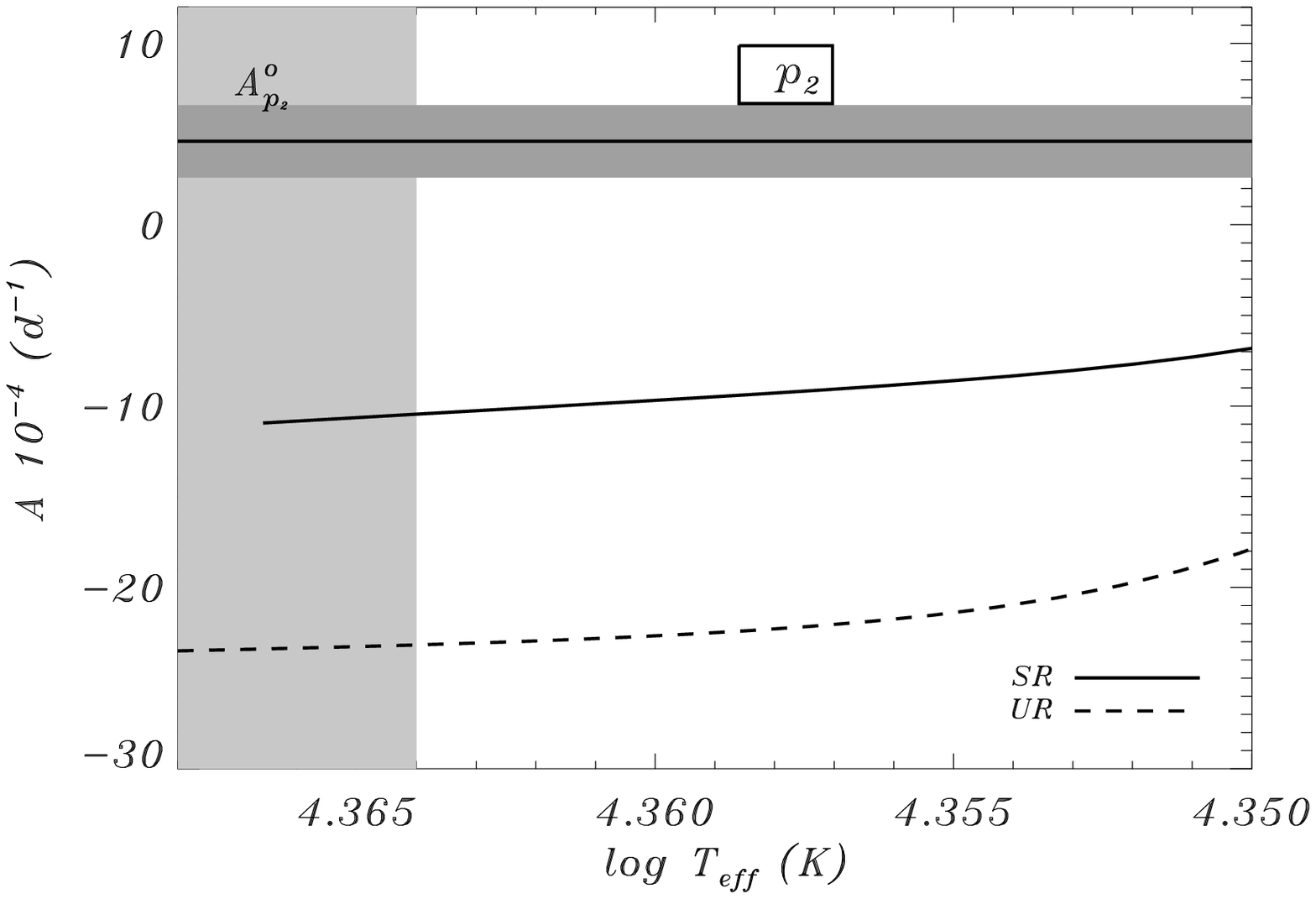}  

    \caption{Evolution of the theoretical semi-splittings (left column) and 
             asymmetries (right column) as a function of the
             effective temperature (logarithmic scale) for the 
	     selected $7.13\,\msol$ model with a rotational velocity
	     of $7\,\kms$ , approximately. 
	     The observed values are depicted with horizontal lines,
	     and the corresponding observational uncertainties are
	     represented by horizontal shaded bands.
	     Each panel row shows the results, from top to bottom,
	     for the $\ell=1$ triplets ${\rm g}_1$, ${\rm p}_1$, and ${\rm
             p}_2$, respectively.
	     In left panels, solid lines represent the positive
	     semi-splittings $\deltap$, and dashed lines, the negative
	     ones $\deltam$. For the asymmetries, the results obtained from UR
             and SR models are represented by dashed and solid lines,
             respectively. The shaded vertical region indicates the range 
             of effective temperatures defined by the models SR1 and UR1 
	     (more details in Section~\ref{ssec:asym}).}
	     \label{fig:deltapmasym}
  \end{center} 
\end{figure*}

\begin{figure}
  \begin{center}
    \includegraphics[width=6.8cm]{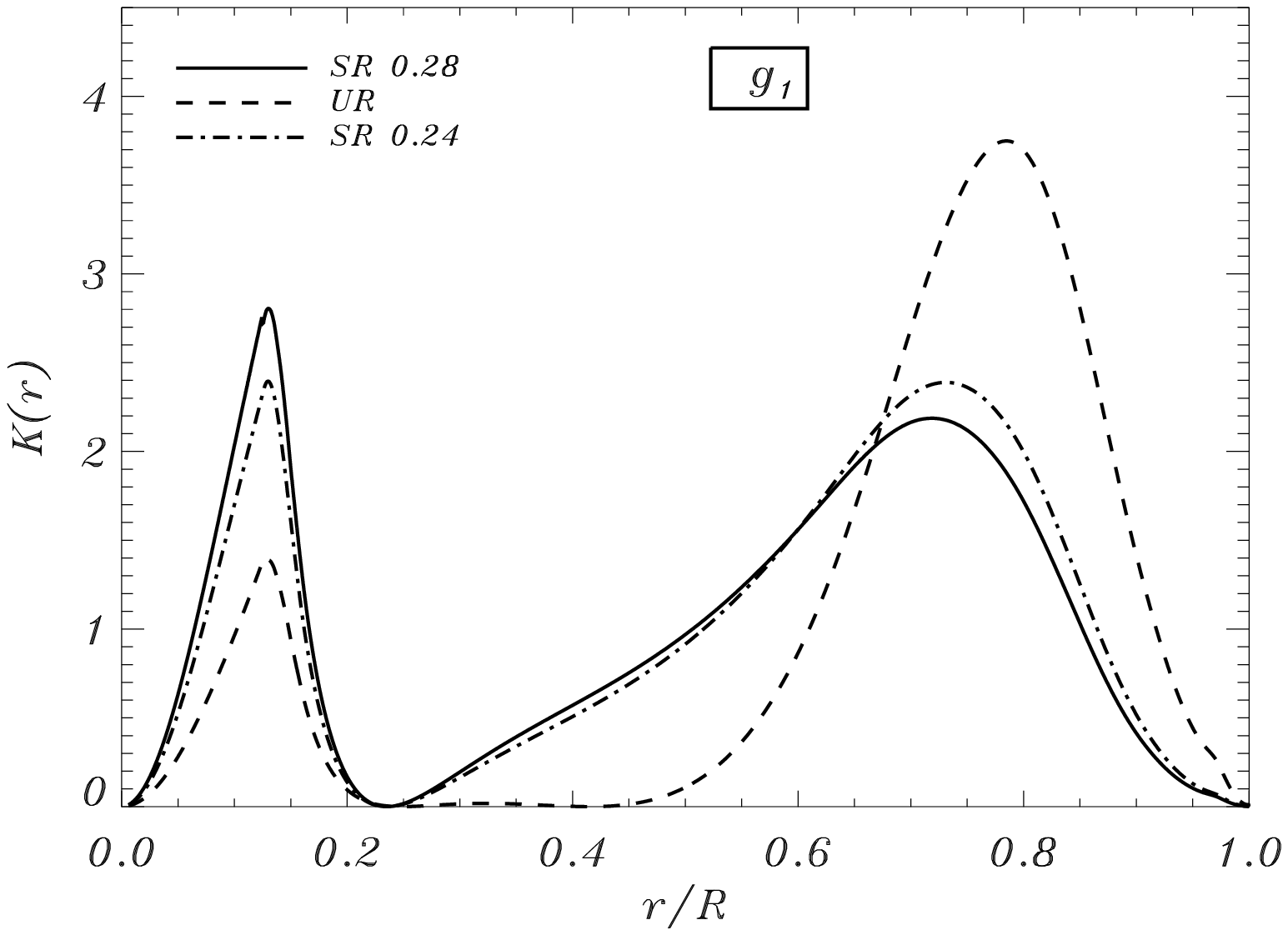}
    \includegraphics[width=6.8cm]{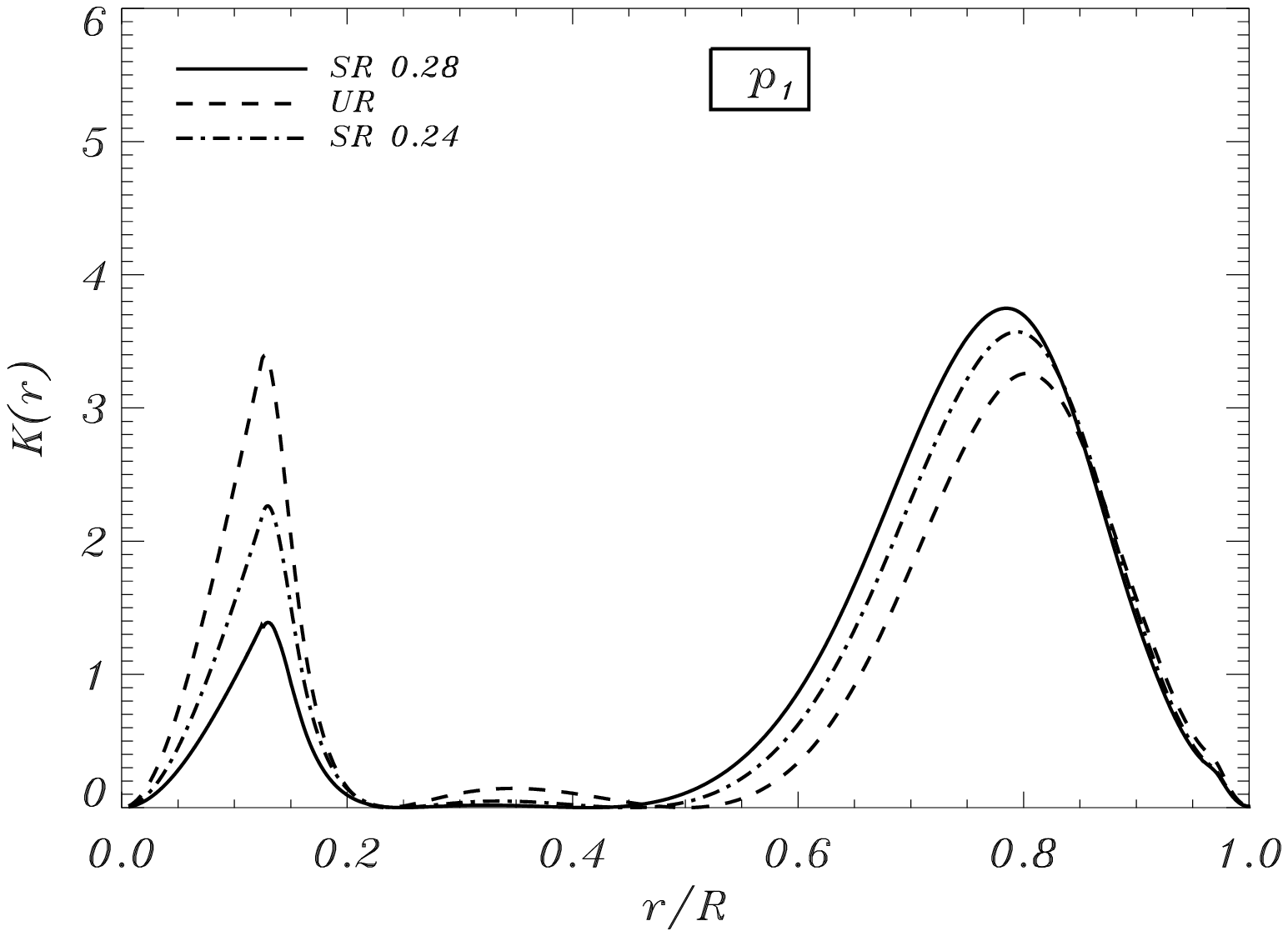}
    \includegraphics[width=6.8cm]{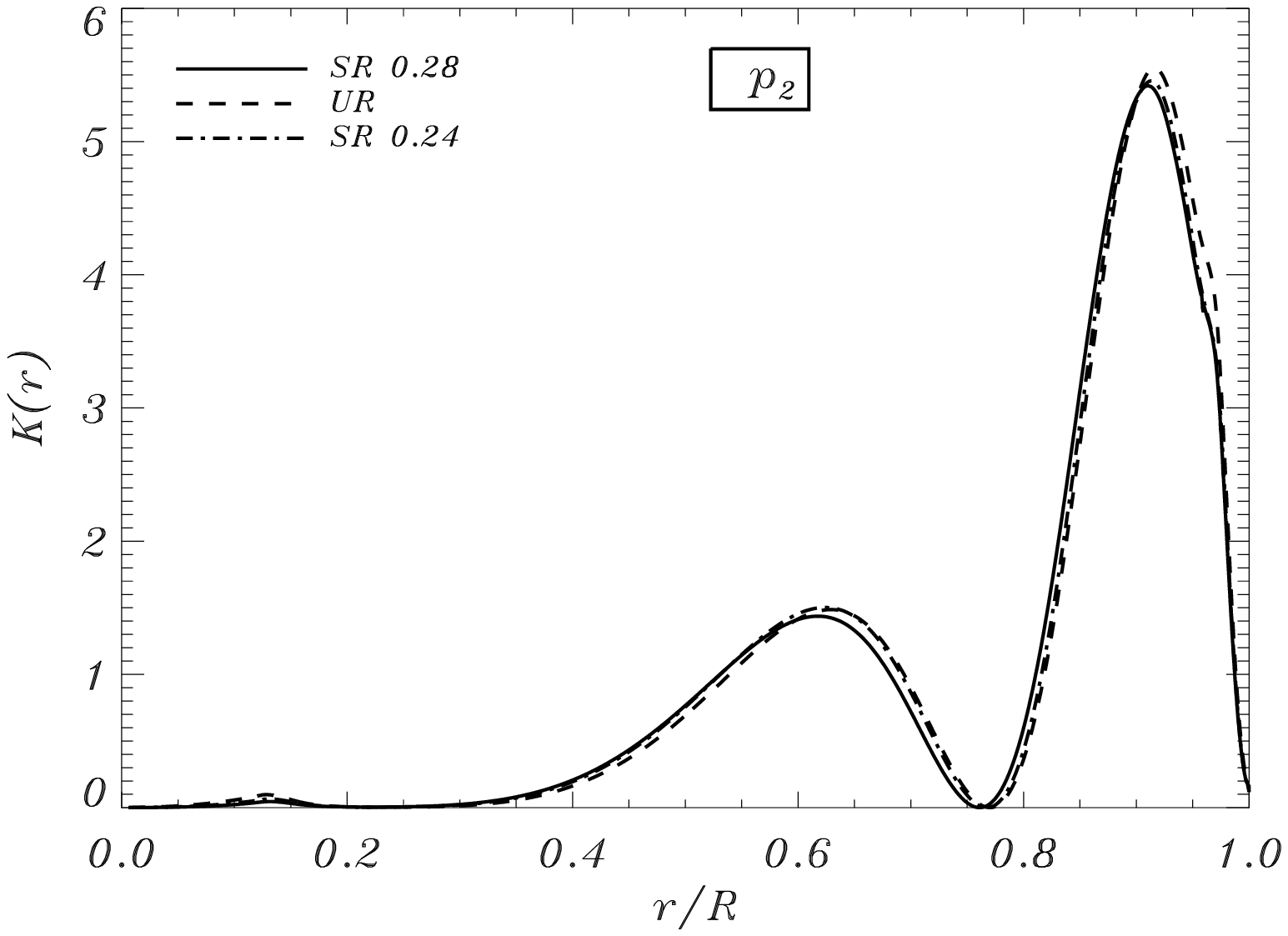}   
    \caption{First-order rotational splitting kernel, from top
             to bottom, for the ${\rm g}_1$, ${\rm p}_1$, and ${\rm p}_2$ modes
             for the
	     selected uniformly-rotating models (UR, dashed line)
	     and differentially-rotating models (SR). For the latter
	     models, two $\dov$ values are considered: 0.28 (continuous
	     line) and 0.24 (dot-dashed line). 
	     Note that the position of the zeros and maxima for UR models 
	     is slightly shifted with respect to those of the SR
	     models. This 
	     can be explained by small differences in
	     the radii of both models, which, as expected, 
	     principally affects to the ${\rm g}_1$ and ${\rm p}_1$ split
             modes. }
	     \label{fig:kernels}
  \end{center} 
\end{figure}

In the case of ${\rm p}_2$, the predicted semi-splittings are larger than
the observed value in the whole range of effective
temperatures, except for $\log\teff\sim4.35$, for which    
SR predictions are almost coincident with the observations
(Fig.~\ref{fig:deltapmasym}, left bottom panel). 
In the shaded region $\edelta=6\times10^{-3}\,\cpd$ for the
SR models and $\edelta=1.6\times10^{-2}\,\cpd$ for the UR
models. Similarly to the ${\rm g}_1$ case, the best results
are given by the SR models for effective temperatures
around the cooler limit of the photometric uncertainty box
($\log\teff=4.35$).
\clearpage
\clearpage
Furthermore, it is worth noting that contrary to the ${\rm p}_1$ and ${\rm g}_1$
cases, the observed semi-splitting $|\deltam|$ is smaller than $|\deltap|$,
whereas both SR and UR models predict $|\deltam|>|\deltap|$
(Fig.~\ref{fig:deltapmasym}, left bottom panel).
This results in a positive observed asymmetry, whereas both UR and SR models
predict negative values (Fig.~\ref{fig:deltapmasym}, right bottom panel).
In the vicinity of $\log\teff=4.365$ (shaded region), the difference between the
observed and predicted asymmetries is about $\adelta=10^{-3}\,\cpd$ 
($\adelta=1.5\times10^{-3}\,\cpd$ for the SR models and 
$\adelta=2.8\times10^{-3}\,\cpd$ for the UR models), which represents a
difference
of one order of magnitude with respect to the observed value. Such apparently
marginal contradiction between predictions and observations could be a
consequence either of 1) an incorrect mode identification, that is, the
observed frequencies concerned do not belong to the rotationally split mode, 
or 2) the use of a wrong description for the rotation profile, particularly
in the outer shells of the star (note the small influence of the selected
rotation profile near the $\mu$-gradient zone, Fig.~\ref{fig:kernels}).
A priori, none of the these possibilities can be discarded. In order to
solve this problem, improvements on both the observations and modelling are
required. From the observational side, an improvement of accuracy with which the
concerned observed frequencies are determined might help to confirm the
observed asymmetry and its sign. Moreover, the detection of additional
frequencies (e.g. with the help of space missions), may provide new insight on
the current mode identification.

From the theoretical side, the second possibility given above is related
with angular momentum redistribution, which plays an important role.
In particular, balance between rotationally induced turbulence and meridional
circulation generates mixing of chemicals and redistribution of angular momentum
\citep{Zahn92}, which affects the rotation profile and the evolution of the
star. The present technique is, therefore, especially suitable for testing that
theory by providing estimates for the coefficients of turbulence using only
asteroseismic observables. This can be illustrated by artificially
modifying the physical conditions beyond the convective core, which can be
done varying the overshooting parameter.

\clearpage
\clearpage

\begin{figure}
  \begin{center}
    \includegraphics[width=8.cm]{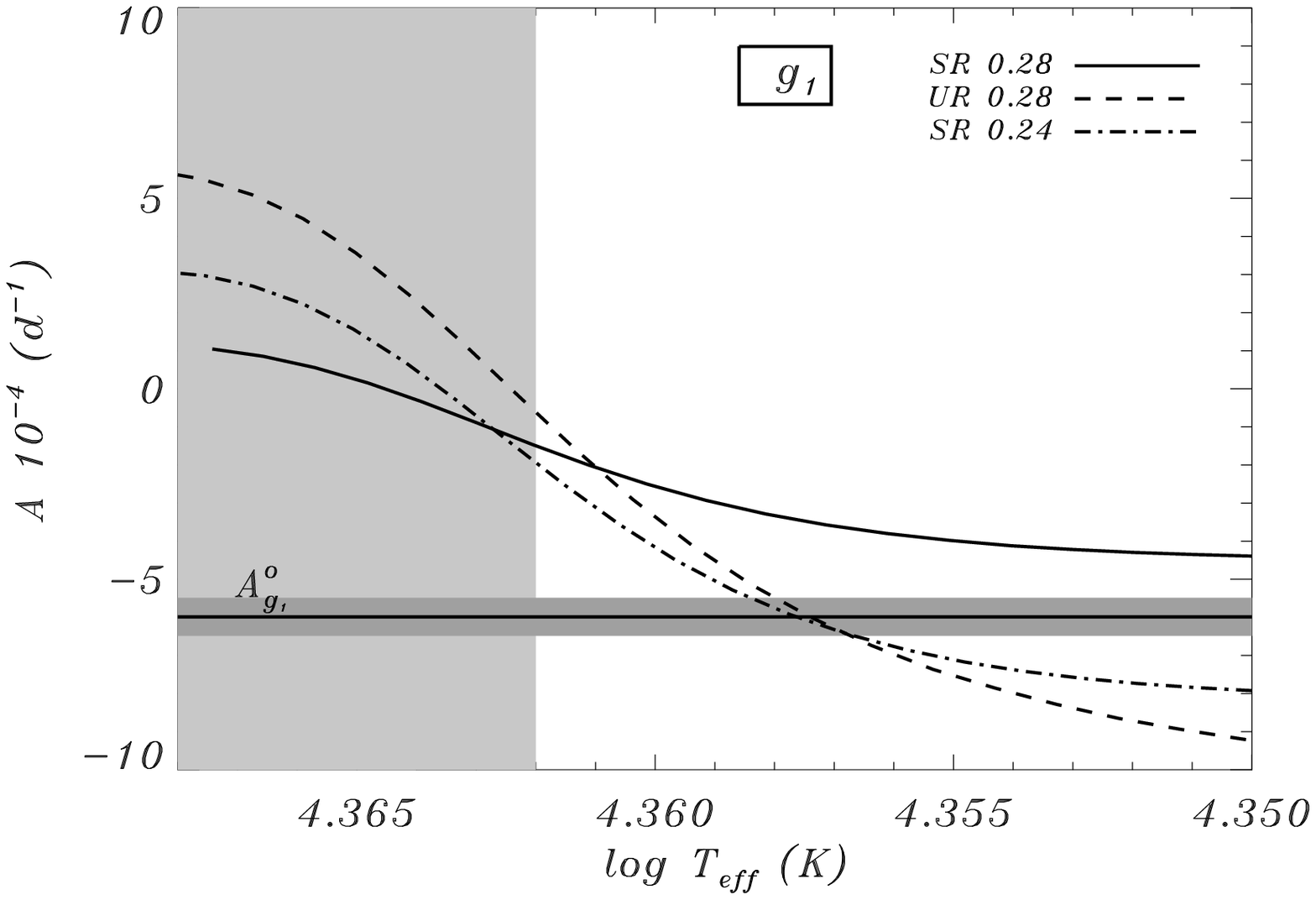}
    \includegraphics[width=8.cm]{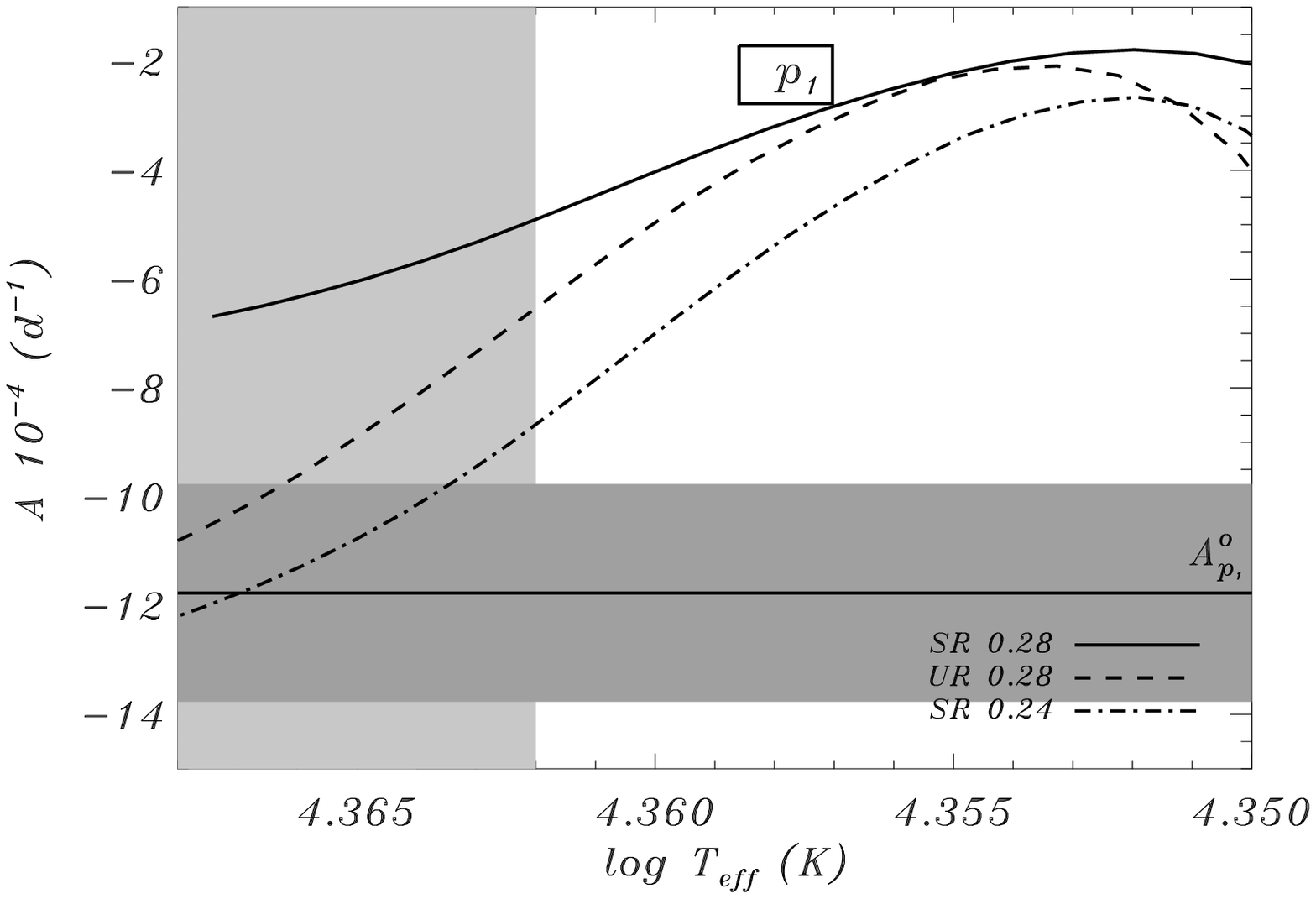}
    \includegraphics[width=8.cm]{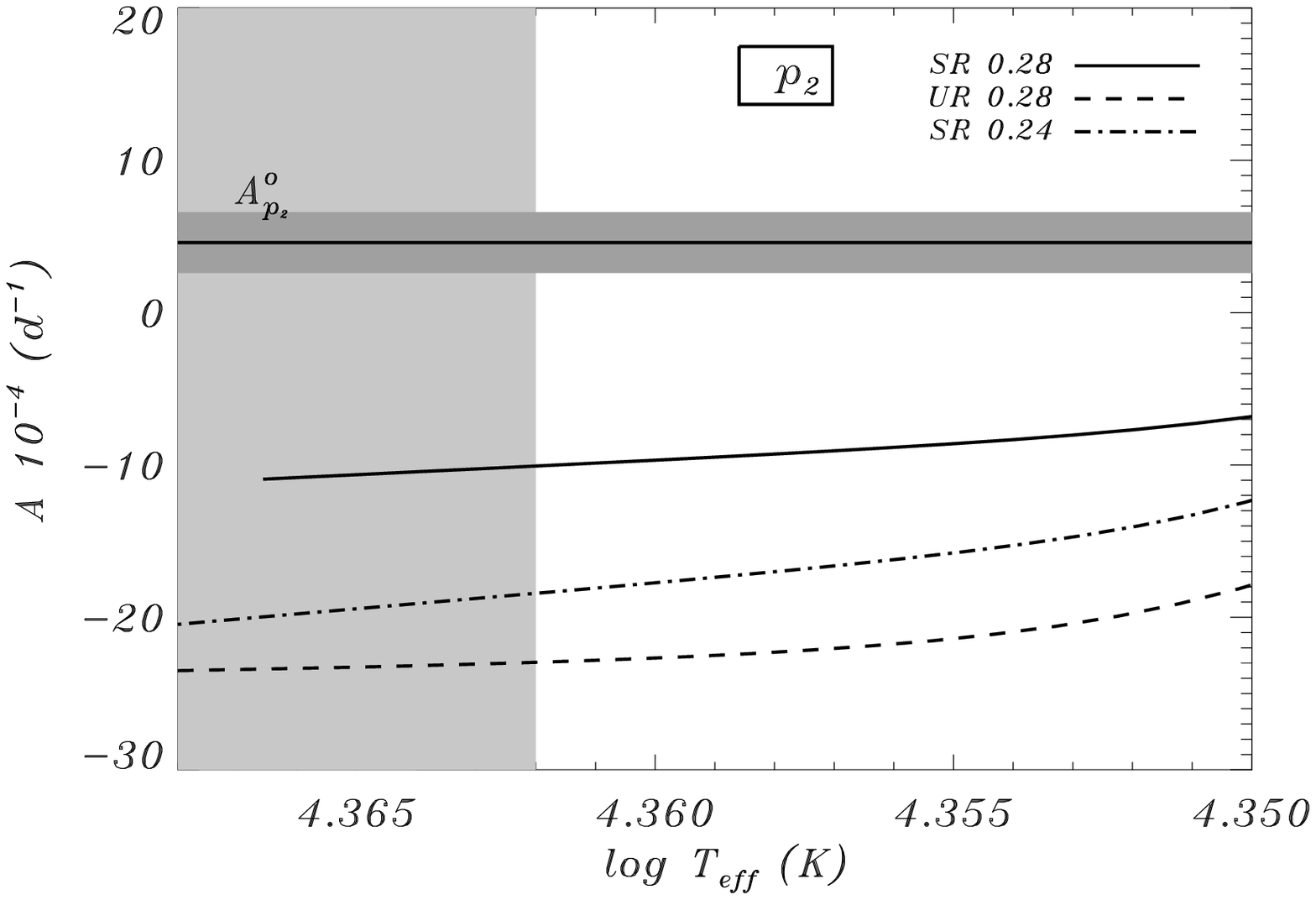} 
    \caption{Comparison of the splitting asymmetries shown in
             Fig.~\ref{fig:evolfreq}
             (for models with $\dov=0.28$), with those given by
	     the models computed with $\dov=0.24$ (dot-dashed line). 
	     The additional vertical shaded band represents the range
	     of $\teff$ defined by the models SR1, SR2, and UR1.}
	     \label{fig:asymcomp}
  \end{center} 
\end{figure}

As expected, such variations principally affect the low-order ${\rm
g}_1$ and ${\rm p}_1$ (see Fig.~\ref{fig:kernels} for a comparison between the
rotational kernels of UR, SR-$\dov=0.28$, and SR-$\dov=0.24$ models).
In Fig.~\ref{fig:asymcomp}, the results for the asymmetries
given in Fig.~\ref{fig:deltapmasym} (right column), which were computed
with an overshooting parameter of $\dov=0.28$, are compared with
those obtained from SR models computed with $\dov=0.24$. 
The best model found for this overshooting parameter value is SR2 (see
Tables~\ref{tab:models} and \ref{tab:freqteor}).
In particular, a variation in $\dov$ of 0.04 results in differences in
the asymmetry of the order of those found between UR and SR ($\dov=0.28$)
models, i.e. a few $10^{-4}\,\cpd$ for ${\rm g}_1$ and ${\rm p}_1$, and about 
$10^{-4}\,\cpd$, which represents almost one order of magnitude smaller
than the difference between the results yield by the UR and SR ($\dov=0.28$)
models. Moreover, in the case of ${\rm p}_1$, the SR ($\dov=0.24$) 
models fit the observed asymmetry in the limits of the observational 
uncertainties. 

\clearpage
\clearpage

Furthermore, recent theoretical studies 
(Andrade, Su\'arez \& Goupil 2008, work in progress), seem
to indicate that even rough variations of the rotation profile may
modify the asymmetries of the split modes. This modification can be as
important as to change the sign of the asymmetries.
According to this, the frequencies around ${\rm p}_2$ cannot be discarded
as belonging to a rotationally split mode. However, if confirmed, 
the presence of such different asymmetries in the oscillation spectrum is
very interesting, since they may significantly constrain  the models
providing information about the structure and rotation profile in the
zones where they have more amplitude.

\section{Conclusions}\label{sec:conclusions}

An asteroseismic  analysis of the  \bceph\ star \nueri\  is presented,
with focus on the study of the internal rotation profile. To this aim,
a new method is presented based on the analysis of rotational
splittings and their asymmetries. Some of the most updated
asteroseismic modelling techniques are used, in particular, analysis
of mode stability and improved descriptions for rotation effects.
Regarding the latter, the so-called pseudo-rotating models are used,
which consider radial differential rotation profiles (shellullar
rotation) in both the evolutionary models and adiabatic oscillations
computations. This represents an important qualitative step with
respect to previous theoretical works.

The present work is divided into two parts. In the first part, a comparison of
the different numerical packages (Li\`ege, Warsaw-New Jersey, and Granada
packages)
was performed (the latter has been used in the present work). This
comparison was performed using a mass-metallicity relation of the models fitting
two of the observed frequencies. In that case, important differences in mass,
up to $0.25\,\msol$ between Li\`ege models and ours, were found. Interestingly,
differences between ASTA04's models and ours can be explained by the
different treatments of the overshooting implemented in the evolutionary codes. 
ASTA04's models only take into account density variations in the overshooted
region, which slightly affects the temperature gradient. On the other hand, our
models were constructed considering an additional restriction by imposing that
the real temperature gradient must equal the adiabatic gradient.
Physically, this implies that, either the heat is transported efficiently
outwards from the stellar core through convective movements due to the
overshooting (the latter case), or purely by radiation (ASTA04's treatment).
This constitutes a very interesting
challenge for asteroseismology, since the oscillation modes (low-order
g and p modes) are sensitive to the physical description of the $\mu$ gradient
zone.

Then, the previous exercise is extended to four frequencies with mode excitation
included. In that case,
models were also constrained to match the fundamental radial mode and three
$m=0$ splitting components. Non-standard models
built with a significant decrease of the initial hydrogen abundance,
$\xin=0.50$, were necessary to match and excite the modes. This provided models
similar, but not identical,\footnote{Cf. the explanation given above when
matching only two frequencies} to those found by ASTA04 using a
solar iron relative abundance, and the same initial hydrogen
abundance. The remaining differences between ASTA04's models and ours indicate
that, for this level of the model accuracy, the present modelling still
depend on the core overshooting. This parameter may change the physical
conditions beyond the convective core. In that region, other physical
processes take place, such as the mixing of chemical elements due to rotation.
In particular, balance between rotationally induced turbulence and meridional
circulation generates mixing of chemicals and redistribution of angular momentum
\citep{Zahn92}. This affects the rotation profile and the evolution of the
star. 

Secondly, the method here presented studies the asymmetries of the split
modes in order to refine the modelling, and more importantly, provide
information about the rotation profile of the star. To do so, pseudo-rotating
models were built using the physical parameters provided in the first part of
the work.
Models with masses around $7.13\,\msol$, and ages around 14.9\,Myr, were found
to fit better 10 of the 14 observed frequencies, which were identified as the
fundamental radial mode and the three $\ell=1$ triplets ${\rm g}_1$, ${\rm
p}_1$, and ${\rm p}_2$. For these modes, a comparison between the observed
and predicted splittings and their asymmetries was performed. Two type of 
rotation profiles were considered: uniform rotation and shellular rotation
profiles. Differences between predictions and observations were found to be of
the order of $5\times10^{-3}$ and $10^{-4}\,\cpd$, for the rotational splittings
and
their asymmetries, respectively. For this last mode, none of the selected models
reproduce neither the splittings, generally larger than the observed ones, nor
the asymmetries, whose predictions have the opposite sign than the observed
values. Although this result might indicate that the frequencies around 
${\rm p}_2, m=0$ mode do not belong to this rotationally-split mode, 
other possibilities cannot be discarded. In fact, a wrong physical description
of the rotation
profile may be responsible for a so peculiar result, particularly in the outer
regions of the star. This result is very important because, up to now, none of
the physical phenomena in which rotation plays a role predict variations
of the rotation profile in that region. Furthermore, it is shown that
asymmetries are quite dependent on the overshooting of the convective core.
Therefore, the method here presented is suitable for testing the theories
describing the angular momentum redistribution and chemical mixing due to
rotationally-induced turbulence.  

In general, the seismic models which include a description for
shellular rotation yield slightly better results as compared with those given
by uniformly-rotating models. Even so, further improvements are necessary in
order to better constrain the modelling of \nueri. In particular, efforts
should be focused in searching a better description of the rotation
profile, for which a detailed study of asymmetries is required (work in
progress). This may be enhanced by analysing other \bceph\ stars with larger
rotational velocities, so that splittings and asymmetries are not of the 
same order than the frequency uncertainties.

\acknowledgments

   JCS acknowledges support by the "Instituto de 
   Astrof\'{\i}sica de Andaluc\'{\i}a" by an I3P contract
   financed by the European Social Fund and from the Spanish 
   "Plan Nacional del Espacio" under project ESP2007-65480-C02-01.
   PJA acknowledges financial support from a "Ramon y Cajal" contract 
   of the Spanish Ministry of Education and Science.
   CRL acknowledges financial support from an "\'Angeles Alvari\~no"
   contract of the "Xunta de Galicia", local government.

\end{document}

%% file: tab1.tex
\begin{deluxetable}{rrc}
  \tablewidth{0pt}
  \tablecaption{A list of independent frequencies
                taken from \citet{Jerzy05nueri}.   \label{tab:freqobs}}
  \tablehead{
       \colhead{ID} & \colhead{$f_{i}$ [$\mathrm{d}^{-1}$]} & 
       \colhead{$f_{i}$ [$\muHz$]} }
      \startdata
       $f_1$ & 5.7632828 & 66.7047 \\               
       $f_2$ & 5.6538767 & 65.4384 \\                    
       $f_3$ & 5.6200186 & 65.0465 \\       
       $f_4$ & 5.6372470 & 65.2459 \\      
       $f_5$ & 7.898200  & 91.4143 \\               
       $f_6$ & 6.243847  & 72.2667 \\                    
       $f_7$ & 6.262917  & 72.4875 \\      
       $f_8$ & 7.20090   & 83.3437 \\ 
       $f_9$ & 7.91383   & 91.5952 \\                    
    $f_{10}$ & 7.92992   & 91.7815 \\      
    $f_{11}$ & 6.73223   & 77.9193 \\ 
    $f_{12}$ & 6.22360   & 72.0324 \\ 
    \enddata
                                  	     
\end{deluxetable}

%% file: tab2.tex
\begin{deluxetable}{lcccccccccccc}

  \tablecolumns{13}
  \tablewidth{0pt}
  \tabletypesize{\small}
  \tablecaption{Characteristics of the best models selected in this work.
	   The different columns are, from left to right: the model
           identification, the metallicity $Z$, the mass $M$, 
	   the radius $R$, the logarithm of the effective temperature $\teff$
           (in K), the logarithm of the gravity $g$ (in cgs), the logarithm of
           the luminosity (cgs), the overshooting parameter (in
           the ${\rm H}_p$ scale), the rotational frequency of the surface
           $\omegas$ (in $\muHz$), the rotational frequency of the core
           $\omegac$ (in $\muHz$), the age (in Myr), the hydrogen abundance in
           the core $\xc$, and the radius of the convective core $\rc$.
           \label{tab:models}}
   \tablehead{   
     \colhead{Model} & \colhead{$Z$} & \colhead{$\mmsol$} & \colhead{$\rrsol$} &
     \colhead{$\lteff$} & \colhead{$\log g$} & \colhead{$\logl$} &
     \colhead{$\dov$} & \colhead{$\omegas$} & \colhead{$\omegac$} &
     \colhead{Age} & \colhead{$\xc$} & \colhead{$\rcr$} }
   \startdata    
        NR1   & 0.019 & 7.13 & 5.704 & 4.364 & 3.778 & 3.923 & 0.28 & 0 & 0 
              & 14.82 & 0.139 & 0.129 \\
        SR1   & 0.019 & 7.13 & 5.714 & 4.364 & 3.777 & 3.923 & 0.28 & 0.258 
              & 0.681 & 14.90 & 0.138 & 0.129  \\
        SR2   & 0.019 & 7.13 & 5.727 & 4.362 & 3.775 & 3.918 & 0.24 & 0.301 
              & 0.778 & 14.60 & 0.129 & 0.125  \\
        UR1   & 0.019 & 7.13 & 5.511 & 4.368 & 3.808 & 3.909 & 0.28 & 0.377 
              & 0.377 & 15.80 & 0.159 & 0.136  \\
   \enddata    
   
\end{deluxetable}

%% file: tab3.tex
\begin{deluxetable}{lcccc}
  \tablewidth{0pt}
  \tablecaption{List of theoretical frequencies ($\cpd$) for the three selected
                best models NR, SR1, SR2, and UR (see Table~\ref{tab:models}). 
	        The model A does not take rotation into account, hence only the
                frequencies of the $m=0$ components are reported.   
                \label{tab:freqteor}}
   \tablehead{  
     \colhead{Mode} & \colhead{$\nu_{{\rm NR},i}$} & 
     \colhead{$\nu_{{\rm SR1},i}$} & \colhead{$\nu_{{\rm SR2},i}$} &
     \colhead{$\nu_{{\rm UR},i}$} }
   \startdata 
    F0               & 5.75818 & 5.74396 & 5.72332 & 5.77681 \\         
    ${\rm g}_{1,-1}$ & -       & 5.65407 & 5.70359 & 5.66402 \\             
    ${\rm g}_{1, 0}$ & 5.63699 & 5.63402 & 5.67979 & 5.64512 \\                
    ${\rm g}_{1,+1}$ & -       & 5.61397 & 5.65585 & 5.62652 \\       
    ${\rm p}_{1,-1}$ & -       & 6.26192 & 6.31067 & 6.27800 \\      
    ${\rm p}_{1, 0}$ & 6.24442 & 6.24026 & 6.28685 & 6.25494 \\             
    ${\rm p}_{1,+1}$ & -       & 6.21785 & 6.26212 & 6.23102 \\                
    ${\rm p}_{2,-1}$ & -       & 7.89759 & 7.87980 & 7.94832 \\      
    ${\rm p}_{2, 0}$ & 7.88829 & 7.87398 & 7.85263 & 7.91898 \\ 
    ${\rm p}_{2,+1}$ & -       & 7.84899 & 7.82361 & 7.88731 \\               
    \enddata    
\end{deluxetable}

%% file: tab4.tex
\begin{deluxetable}{ccccc}
  \tablewidth{0pt}
  \tablecaption{Observed semi-splittings, $(\deltap, \deltam)$, and 
                asymmetries, $A^{\rm o}$ of the $\ell=1$ triplets 
	        ${\rm g}_1$, ${\rm p}_1$ and ${\rm p}_2$, as identified using 
	        a model of $7.13\,\msol$ which rotates with 
	        $v\sim5.5\,\kms$ in the surface. Quantities are 
	        given in $\cpd$. Uncertainties, $\delta$, are calculated
	        from data in \citet{Jerzy05nueri}.\label{tab:identif}}
   \tablehead{  
     \colhead{Mode}        & 
     \colhead{$\deltap$}   & 
     \colhead{$\deltam$}   &
     \colhead{$A^{\rm o}$} \\
      &
     \colhead{$\delta(\deltap)$} &
     \colhead{$\delta(\deltam)$} & 
     \colhead{$\delta(A^{\rm o})$} }   
   \startdata               
      ${\rm g}_1$ & $0.016630$  & $-0.01723$ & $-0.000598$ \\ 
       $ $    & $\pm6\times10^{-6}$ & $\pm7\times10^{-6}$ & $\pm1\times10^{-5}$  \\     
      ${\rm p}_1$ & $0.0191$    & $-0.0203$  & $-0.0012$   \\  
       $ $     & $\pm2\times10^{-4}$ & $\pm2\times10^{-4}$ & $\pm2\times10^{-4}$  \\ 	
      ${\rm p}_2$ & $0.0161$    & $-0.0156$  & $+0.0005$   \\
       $ $ & $\pm1\times10^{-4}$ & $\pm2\times10^{-4}$ & $\pm2\times10^{-4}$  \\  
    \enddata    
\end{deluxetable}

%% file: ms_astroph.bbl
\begin{thebibliography}{38}

\bibitem[{{Aerts} {et~al.}(2004){Aerts}, {De Cat}, {Handler}, {Heiter},
  {Balona}, {Krzesinski}, {Mathias}, {Lehmann}, {Ilyin}, {De Ridder},
  {Dreizler}, {Bruch}, {Traulsen}, {Hoffmann}, {James}, {Romero-Colmenero},
  {Maas}, {Groenewegen}, {Telting}, {Uytterhoeven}, {Koen}, {Cottrell},
  {Bentley}, {Wright}, \& {Cuypers}}]{Aerts04nueri}
{Aerts}, C., {De Cat}, P., {Handler}, G., {et~al.} 2004, \mnras, 347, 463

\bibitem[{{Alexander} \& {Ferguson}(1994)}]{AlexFergu94}
{Alexander}, D.~R. \& {Ferguson}, J.~W. 1994, \apj, 437, 879

\bibitem[{{Ausseloos} {et~al.}(2004){Ausseloos}, {Scuflaire}, {Thoul}, \&
  {Aerts}}]{Ausseloos04}
{Ausseloos}, M., {Scuflaire}, R., {Thoul}, A., \& {Aerts}, C. 2004, \mnras,
  355, 352 (ASTA04)
  
\bibitem[{{Baglin} {et~al.}(2002){Baglin}, {Auvergne}, {Barge}, {Buey},
  {Catala}, {Michel}, {Weiss}, \& {COROT Team}}]{Baglin02}
{Baglin} A., {Auvergne} M., {Barge} P., {Buey} J.-T., {Catala} C., {Michel} E.,
  {Weiss} W., {COROT Team}, 2002, in {Battrick} B., {Favata} F., {Roxburgh}
I.~W.,
  {Galadi} D., eds, ASP Conf. Ser. Vol.~259, Radial and Nonradial
  Pulsations as Probes of Stellar Physics. Astron. Soc. Pac.,
  San Francisco, p.~17    
  
\bibitem[{{B{\" o}hm-Vitense}(1958)}]{BohmVit58}
{B{\" o}hm-Vitense} E., 1958, Zeitschrift f\"ur Astrophysics, 46, 108
  
\bibitem[{{Boury} {et~al.}(1975){Boury}, {Gabriel}, {Noels}, {Scuflaire},\&
{Ledoux}}]{Boury75cles}
{Boury}, A. and {Gabriel}, M. and {Noels}, A. and {Scuflaire}, R. and 
{Ledoux}, P. 1975, \aap, 41, 279

\bibitem[{{Christensen-Dalsgaard} \& {Daeppen}(1992)}]{ceff}
{Christensen-Dalsgaard}, J. \& {Daeppen}, W. 1992, \aapr, 4, 267

\bibitem[{{Clayton}(1968)}]{Clayton68}
{Clayton}, D.~D. 1968, {Principles of stellar evolution and nucleosynthesis}
  (New York: McGraw-Hill, 1968)

\bibitem[{{De Ridder} {et~al.}(2004){De Ridder}, {Telting}, {Balona},
  {Handler}, {Briquet}, {Daszy{\' n}ska-Daszkiewicz}, {Lefever}, {Korn},
  {Heiter}, \& {Aerts}}]{DeRidder04}
{De Ridder}, J., {Telting}, J.~H., {Balona}, L.~A., {et~al.} 2004, \mnras, 351,
  324

\bibitem[{{Dupret} {et~al.}(2002){Dupret}, {De Ridder}, {Neuforge}, {Aerts}, \&
  {Scuflaire}}]{Dupret02}
{Dupret}, M.-A., {De Ridder}, J., {Neuforge}, C., {Aerts}, C., \& {Scuflaire},
  R. 2002, \aap, 385, 563

\bibitem[{{Dziembowski} \& {Jerzykiewicz}(2003)}]{Dziembowski03nueri}
{Dziembowski}, W.~A. \& {Jerzykiewicz}, M. 2003, in Astronomical Society of the
  Pacific Conference Series, 319

\bibitem[{{Dziembowski} \& {Pamiatnykh}(1993)}]{DziembowskiAlosha93}
{Dziembowski}, W.~A. \& {Pamiatnykh}, A.~A. 1993, \mnras, 262, 204

\bibitem[{{Eggleton} {et~al.}(1973){Eggleton}, {Faulkner}, \&
  {Flannery}}]{Eggleton73}
{Eggleton}, P.~P., {Faulkner}, J., \& {Flannery}, B.~P. 1973, \aap, 23, 325

\bibitem[{{Gautschy} \& {Saio}(1993)}]{GautschySaio93}
{Gautschy}, A. \& {Saio}, H. 1993, \mnras, 262, 213

\bibitem[{{Grevesse} \& {Noels}(1996)}]{GrevesseNoels96}
{Grevesse}, N. \& {Noels}, A. 1996, in Holt, S.~S. and Sonneborn, G. eds,
Vol.~99, Astron. Soc. Pac. Conference Series, 117

\bibitem[{{Handler} {et~al.}(2004){Handler}, {Shobbrook}, {Jerzykiewicz},
  {Krisciunas}, {Tshenye}, {Rodr{\'{\i}}guez}, {Costa}, {Zhou}, {Medupe},
  {Phorah}, {Garrido}, {Amado}, {Papar{\' o}}, {Zsuffa}, {Ramokgali}, {Crowe},
  {Purves}, {Avila}, {Knight}, {Brassfield}, {Kilmartin}, \&
  {Cottrell}}]{Handler04nueri}
{Handler}, G., {Shobbrook}, R.~R., {Jerzykiewicz}, M., {et~al.} 2004, \mnras,
  347, 454

\bibitem[{{Iglesias} \& {Rogers}(1996)}]{Igle96}
{Iglesias}, C.~A. \& {Rogers}, F.~J. 1996, \apj, 464, 943

\bibitem[{{Jerzykiewicz},  {et~al.}(2005){Jerzykiewicz}, {Handler}, {Shobbrook},
{Pigulski}, {Medupe}, 
         {Mokgwetsi}, {Tlhagwane},\& {Rodr{\'{\i}}guez}}]{Jerzy05nueri}
	 {Jerzykiewicz}, M., {Handler}, G. \& {Shobbrook}, R.~R., {et~al.} 2005,
\mnras,
         360, 619
\bibitem[{{Kippenhahn} \& {Weigert}(1990)}]{KipWeig90}
{Kippenhahn}, R. \& {Weigert}, A. 1990, "Stellar structure and evolution",
  Astronomy and Astrophysics library (Springer-Verlag)

\bibitem[{{Kurucz}(1993)}]{Kurucz93cd13}
{Kurucz}, R. 1993, ATLAS9 Stellar Atmosphere Programs and 2 km/s grid.~Kurucz
  CD-ROM No.~13.~ Cambridge, Mass.: Smithsonian Astrophysical Observatory,
  1993., 13
  
\bibitem[{{Maeder} \& {Meynet}(2004)}]{MaederMeynet04}
{Maeder}, A. \& {Meynet}, G. 2004, in Maeder, A. and Eenens, P. eds,
IAU Symposium, Vol.~215, Evolution of Massive Stars with Rotation and Mass Loss.
Astron. Soc. Pac., San Franciscop.~500

\bibitem[{{Michel} {et~al.}(1999){Michel},{Hern{\' a}ndez}, {Houdek}, 
        {Goupil},{Lebreton}, {Hern{\' a}ndez},
        {Baglin}, {Belmonte}, {Soufi}}]{MiHer99}
	{Michel}, E. and {Hern{\' a}ndez}, M.~M. and {Houdek}, G. and
        {Goupil}, M.~J. and {Lebreton}, Y. and {Hern{\' a}ndez}, F.~P. and
        {Baglin}, A. and {Belmonte}, J.~A. and {Soufi}, F. 1999, \aap, 342, 153 
  		
\bibitem[{{Morel}(1997)}]{Morel97}
{Morel}, P. 1997, \aaps, 124, 597

\bibitem[{{Morel} {et~al.}(2006){Morel}, {Butler}, {Aerts}, {Neiner}, 
\& {Briquet}}]{MorelT06}
{Morel}, T. and {Butler}, K. and {Aerts}, C. and {Neiner}, C. and 
{Briquet}, M. 2006, \aap, 457, 651

\bibitem[{{Moya} {et~al.}(2004){Moya}, {Garrido}, \& {Dupret}}]{Moya04}
         {Moya}, A. and {Garrido}, R. and {Dupret}, M.~A. 2004, \aap, 414, 1081
	 
\bibitem[{{Moya} {et~al.}(2007){Moya},\& {Garrido}}]{Moya08graco}
         {Moya}, A. and {Garrido}, R. 2008, \apss, in press.	 
	 
\bibitem[{{Moya} {et~al.}(2008){Moya}, {Christensen-Dalsgaard}, {Charpinet},
	{Lebreton}, {Miglio}, {Montalban}, {Monteiro},  
	{Provost}, {Roxburgh}, {Scuflaire}, 
	{Suarez},  \& {Suran}}]{Moya08estacorot}
	{Moya}, A. and {Christensen-Dalsgaard}, J. and {Charpinet}, S. and 
	{Lebreton}, Y. and {Miglio}, A. and {Montalban}, J. and {Monteiro},
M.~J.~P.~F.~G. and 
	{Provost}, J. and {Roxburgh}, I.~W. and {Scuflaire}, R. and 
	{Su\'arez}, J.~C. and {Suran}, M. 2008, \apss, in press.

		
\bibitem[{{Pamyatnykh}(1975)}]{Alosha75}
{Pamiatnykh} A.~A., 1975, in {Sherwood}, V.~E., {Plaut}, L. eds, IAU Symp. 67.
 Variable Stars and Stellar Evolution. Dordrecht, Reidel Publishing, pp.~247	
 
\bibitem[{{Pamyatnykh} {et~al.}(2004){Pamyatnykh}, {Handler}, \&
  {Dziembowski}}]{Alosha04nueri}
{Pamyatnykh}, A.~A., {Handler}, G., \& {Dziembowski}, W.~A. 2004, \mnras, 350,
  1022
 
\bibitem[{{P{\'e}rez Hern{\'a}ndez} {et~al.}(1999){P{\'e}rez Hern{\'a}ndez},  
          {Claret},{Hern{\'a}ndez}, \& {Michel}}]{Pe99}
	  {P{\'e}rez Hern{\'a}ndez}, F. and {Claret}, A. and {Hern{\'a}ndez},
M.\ M. and
          {Michel}, E. 1999, \aap, 346, 586
	  
\bibitem[{{Schnerr} {et~al.}(2006){Schnerr}, {Verdugo}, {Henrichs}, 
 \& {Neiner}}]{Schnerr06}
{Schnerr}, R.~S. and {Verdugo}, E. and {Henrichs}, H.~F. 
and {Neiner}, C. 2006, \aap, 452, 969

\bibitem[{{Scuflaire} {et~al.}(2007){Scuflaire},{Th{\'e}ado},{Montalb{\'a}n},
{Miglio}, {Bourge}, {Godart}, {Thoul}, \& {Noels}}]{Scuflaire07}
{Scuflaire}, R. and {Th{\'e}ado}, S. and {Montalb{\'a}n}, J. and 
{Miglio}, A. and {Bourge}, P.-O. and {Godart}, M. and {Thoul}, A. and 
{Noels}, A. 2007, (\apss in press), arXiv:0712.3471

\bibitem[{{Smith}, M.~A.(1983){Smith}}]{Smith83}
         {Smith}, M.~A. 1983, \apj, 265, 338

\bibitem[{{Su{\' a}rez}(2002)}]{SuaThesis}
{Su{\' a}rez}, J.~C. 2002, Ph.D.~Thesis, ISBN 84-689-3851-3, ID 02/PA07/7178

\bibitem[{{Su{\'a}rez} {et~al.}(2006){Su{\'a}rez}, {Goupil}, \&
  {Morel}}]{Sua06rotcel}
{Su{\'a}rez}, J.~C., {Goupil}, M.~J., \& {Morel}, P. 2006, \aap, 449, 673

\bibitem[{{Su\'arez} {et~al.}(2007){Su\'arez},\& {Goupil}}]{Sua08filou}
         {Su\'arez}, J.~C. and {Goupil}, M.~J. 2008, \apss, in press	
	  
\bibitem[{{Su{\'a}rez} {et~al.}(2007){Su{\'a}rez}, {Michel}, {Houdek}, 
         {P{\'e}rez Hern{\'a}ndez}, F. \& {Lebreton}}]{Sua07gammes}
	 {Su{\'a}rez}, J.~C. and {Michel}, E. and {Houdek}, G. and {P{\'e}rez
Hern{\'a}ndez}, F. and 
	 {Lebreton}, Y. 2007, \mnras, 379, 201	 

\bibitem[{{Tran Minh} \& {L\'eon}(1995)}]{filou}
{Tran Minh}, F. \& {L�on}, L. 1995, in Roxburg, I.~W., Maxnou, J.~L., eds.,
  Physical Processes in Astrophysics. Springer Verlag, Berlin., 219

\bibitem[{{Unno} {et~al.}(1989){Unno}, {Osaki}, {Ando}, {Saio}, \&
  {Shibahashi}}]{Unno89}
{Unno}, W., {Osaki}, Y., {Ando}, H., {Saio}, H., \& {Shibahashi}, H. 1989,
  {Nonradial oscillations of stars} (Nonradial oscillations of stars, Tokyo:
  University of Tokyo Press, 1989, 2nd ed.)
  
\bibitem[{{van Hoof}, A.(1961){van Hoof}}]{vanHoof61}
       {van Hoof}, A. 1961, Zeitschrift fur Astrophysik, 53, 106
       
\bibitem[{{Dziembowski} \& {Pamyatnykh}(2008){Dziembowski},  
 \& {Pamyatnykh}}]{Wojtek08nueri} 
  {Dziembowski}, W.~A. and {Pamyatnykh}, A.~A. 2008, \mnras, 385, 2061
  
\bibitem[{{Zahn}(1992)}]{Zahn92}
{Zahn} J.-P., 1992, \aap, 265, 115  

\end{thebibliography}
